\documentclass[a4paper]{article}
\pdfoutput=1
\usepackage{jheppub}
\usepackage{booktabs}
\usepackage{multirow}
\usepackage{graphicx}
\usepackage{epsfig}
\usepackage{lscape}
\usepackage{rotating}
\usepackage{hyperref}
\usepackage{amsmath}
\usepackage{lineno}
\usepackage{hyperref}
\usepackage{graphicx}
\usepackage{hyperref}
\usepackage{natbib}
\usepackage{subcaption}
\usepackage{epstopdf}
\usepackage{url}
\usepackage[utf8]{inputenc}
\usepackage{amsmath,amssymb,verbatim,mathrsfs}
\usepackage{textcomp}

\graphicspath{{Figs/}}

\title{Review of LHC experimental results on low mass bosons in multi Higgs models}

\author[1,2,3]{R. Aggleton}
\author[4]{D. Barducci}
\author[5]{N-E. Bomark}
\author[1,2]{S. Moretti}
\author[1]{C. Shepherd-Themistocleous}

\emailAdd{robin.aggleton@bristol.ac.uk}
\emailAdd{barducci@lapth.cnrs.fr}
\emailAdd{nilserik.bomark@gmail.com}
\emailAdd{s.moretti@soton.ac.uk}
\emailAdd{claire.shepherd@stfc.ac.uk}

\affiliation[1]{Particle Physics Department, Rutherford Appleton Laboratory, Chilton, Didcot, Oxon OX11 0QX, UK}
\affiliation[2]{School of Physics \& Astronomy, University of Southampton, Highfield, Southampton SO17 1BJ, UK}
\affiliation[3]{H. H. Wills Physics Laboratory, Bristol University, Bristol BS8 1TH, UK}
\affiliation[4]{LAPTh, Universit\'e Savoie Mont Blanc, CNRS B.P. 110, F-74941 Annecy-le-Vieux, France}
\affiliation[5]{Department of Natural Sciences, University of Agder, Postboks 422, 4604 Kristiansand, Norway}

\begin{document}

\mathchardef\mhyphen="2D  

\newcommand{\invis}{\ensuremath{\mathrm{Invis.}}}
\newcommand{\ttbar}{\ensuremath{t\bar{t}}}
\newcommand{\qqbar}{\ensuremath{q\bar{q}}}
\newcommand{\bbbar}{\ensuremath{b\bar{b}}}
\newcommand{\ccbar}{\ensuremath{c\bar{c}}}
\newcommand{\ssbar}{\ensuremath{s\bar{s}}}
\newcommand{\tauh}{\ensuremath{\tau_{h}}}
\newcommand{\pp}{\ensuremath{pp}}
\newcommand{\egamma}{\ensuremath{e/\gamma}}
\newcommand{\hsm}{\ensuremath{h_{SM}}}
\newcommand{\hdisc}{\ensuremath{h_{125}}}
\newcommand{\hone}{\ensuremath{h_{1}}}
\newcommand{\htwo}{\ensuremath{h_{2}}}
\newcommand{\hthree}{\ensuremath{h_{3}}}
\newcommand{\hc}{\ensuremath{h^{\pm}}}
\newcommand{\aone}{\ensuremath{a_{1}}}
\newcommand{\atwo}{\ensuremath{a_{2}}}
\newcommand{\phione}{\ensuremath{\phi_{1}}}
\newcommand{\twotautwomu}{\ensuremath{2\tau 2\mu}}
\newcommand{\fourtau}{\ensuremath{4\tau}}
\newcommand{\tautau}{\ensuremath{\tau\tau}}
\newcommand{\mumu}{\ensuremath{\mu\mu}}
\newcommand{\fourmu}{\ensuremath{4\mu}}
\newcommand{\diphoton}{\ensuremath{\gamma\gamma}}
\newcommand{\twobtwomu}{\ensuremath{2 b 2\mu}}

\newcommand{\mh}{\ensuremath{m_h}}
\newcommand{\mchi}{\ensuremath{m_{\chi}}}
\newcommand{\maone}{\ensuremath{m_{\aone}}}
\newcommand{\mhone}{\ensuremath{m_{\hone}}}
\newcommand{\mhtwo}{\ensuremath{m_{\htwo}}}
\newcommand{\mhc}{\ensuremath{m_{\hc}}}
\newcommand{\mphione}{\ensuremath{m_{\phione}}}
\newcommand{\mell}{\ensuremath{m_{\ell}}}
\newcommand{\mmu}{\ensuremath{m_{\mu}}}
\newcommand{\mtau}{\ensuremath{m_{\tau}}}
\newcommand{\mb}{\ensuremath{m_{b}}}
\newcommand{\mqbar}{\ensuremath{\bar{m}_{q}}}

\newcommand{\GeV}{\ensuremath{\mathrm{GeV}}}
\newcommand{\TeV}{\ensuremath{\mathrm{TeV}}}

\newcommand{\cm}{\ensuremath{\mathrm{cm}}}

\newcommand{\pb}{\ensuremath{\mathrm{pb}}}
\newcommand{\fb}{\ensuremath{\mathrm{fb}}}

\newcommand{\invfb}{\ensuremath{\mathrm{fb}^{-1}}}

\newcommand{\MET}{\ensuremath{\slashed{E}_T}}
\newcommand{\METNOMU}{\ensuremath{\slashed{E}^{\mathrm{no}\mhyphen\mu}_T}}
\newcommand{\METNOE}{\ensuremath{\slashed{E}^{\mathrm{no}\mhyphen e}_T}}
\newcommand{\PT}{\ensuremath{p_T}}
\newcommand{\ET}{\ensuremath{E_T}}
\newcommand{\ETA}{\ensuremath{\eta}}
\newcommand{\DETA}{\ensuremath{\Delta\eta}}
\newcommand{\DETAJJ}{\ensuremath{\Delta\eta_{jj}}}
\newcommand{\Mjj}{\ensuremath{M_{jj}}}
\newcommand{\abseta}{\ensuremath{|\eta|}}
\newcommand{\DPHI}{\ensuremath{\Delta\phi}}
\newcommand{\DPHIJJ}{\ensuremath{\Delta\phi_{jj}}}
\newcommand{\DR}{\ensuremath{\Delta R}}
\newcommand{\MT}{\ensuremath{M_{T}}}
\newcommand{\sqrts}{\ensuremath{\sqrt{s}}}
\newcommand{\mmutrack}{\ensuremath{m_{\mu \mhyphen \mathrm{trk}}}}
\newcommand{\kT}{\ensuremath{\mathrm{k}_{\mathrm{T}}}}
\newcommand{\chisq}{\ensuremath{\chi^2}}
\newcommand{\amm}{\ensuremath{a_{\mu}}}
\newcommand{\Delamu}{\ensuremath{\Delta a_{\mu}}}
\newcommand{\dmrelic}{\ensuremath{\Omega_{\mathrm{DM}}h^2}}
\newcommand{\rd}{\ensuremath{R(D)}}
\newcommand{\rds}{\ensuremath{R(D^*)}}
\newcommand{\alphas}{\ensuremath{\alpha_{s}}}
\newcommand{\alphasbar}{\ensuremath{\bar{\alpha}_s}}

\newcommand{\nt}{\texttt{NMSSMTools}}
\newcommand{\hb}{\texttt{HiggsBounds}}
\newcommand{\hs}{\texttt{HiggsSignals}}
\newcommand{\lilith}{\texttt{Lilith}}
\newcommand{\micromegas}{\texttt{micrOMEGAs}}
\newcommand{\superiso}{\texttt{superiso}}

\newcommand{\mueff}{\ensuremath{\mu_{\rm{eff}}}}

\newcommand{\lam}{\ensuremath{{\lambda}}}
\newcommand{\kap}{\ensuremath{{\kappa}}}
\newcommand{\tanbeta}{\ensuremath{\tan\beta}}

\newcommand{\alam}{\ensuremath{{A_{\lambda}}}}
\newcommand{\akap}{\ensuremath{{A_{\kappa}}}}
\newcommand{\mhalf}{\ensuremath{ m_{1/2}}}

\newcommand{\ie}{i.e.\ }
\newcommand{\eg}{e.g.\ }

\abstract{A number of searches at the LHC looking for low mass ($2\mmu$ -- $62\ \GeV$) bosons in $\sqrts = 8\ \TeV$ data have recently been published. We summarise the most pertinent ones, and look at how their limits affect a variety of supersymmetric and non-supersymmetric models which can give rise to such light bosons: the 2HDM (Types I and II), the NMSSM, and the nMSSM.}

\date\today
\begin{flushright}
    \hspace{3cm}LAPTH-040/16 \\
\end{flushright}

\maketitle
\section{Introduction and Motivation}
\label{sec:introduction}

Since the discovery of a Higgs boson in July 2012 by  the ATLAS and CMS collaborations at the Large Hadron Collider (LHC)  \cite{Aad:2012tfa,Chatrchyan:2012xdj} innumerable analyses have been performed in order to ascertain its nature.
While its profile is largely consistent with the predictions of the Standard Model (SM), there remains the possibility that this
object belongs to a Beyond the SM (BSM) scenario in which a SM-like Higgs state is realised in specific configurations of the corresponding parameter space. Since the necessity of BSM physics is evident from both the theoretical
(hierarchy problem, absence of coupling unification, etc.) and experimental  (neutrino masses, dark matter, etc.) point of view,
it is of paramount importance to investigate whether it is possible to access it through Higgs analyses.

A possibility is clearly to improve the precision of the measurements of the discovered SM-like objects as, sooner or later,
statistically significant deviations from the SM predictions may well appear. It should be emphasised, however, that accessing BSM physics indirectly, i.e., through the study of SM-like production and decay channels of the 125 GeV Higgs boson, may not be the most efficient way of isolating the underlying BSM scenario.

An alternative procedure is the following one. Whichever the BSM scenario encompassing the discovered SM-like Higgs state,
this obviously includes an extended Higgs sector, with respect to the SM, hence a Higgs mechanism of Electroweak Symmetry
Breaking (EWSB) giving rise to more physical Higgs states than just the single one of the SM. Crucially, other than
with SM states, all these  emerging Higgs boson (both neutral and charged, both scalar and pseudoscalar), can interact with each other. For example,
the heavier Higgs states can decay into the lighter ones and in these chains the 125 GeV Higgs boson could, if appearing, either be the initiator or else the end product of the various possible decay patterns. Needless to say, to isolate one or more of the latter would be a direct evidence of a non-SM Higgs sector, hence of the existence of BSM physics. Furthermore, the study of the additional Higgs states would certainly gain one much more understanding of the underlying scenario than what can be extracted from the aforementioned analyses of the SM-like Higgs state.

It is the purpose of this paper to review  both the theoretical and experimental status of several BSM scenarios predicting such Higgs cascade decays, in particular, those embedding in their particle spectrum a rather light state, with mass
below 60 GeV or so, which would be produced in pairs in the last step of the  discussed Higgs cascade decays.
From the theoretical side, we will concentrate on the most popular BSM Higgs scenarios in which such a light object is realised,
which is typically pseudoscalar in nature. From the experimental side, we will
adopt published data obtained by the end of Run 1 of the LHC from either ATLAS and CMS, covering several
signatures of such a pair of pseudoscalar Higgs states, including decays into pairs of muons, taus, and bottom quarks.

It is natural to organise the discussion of the possible BSM scenarios behind such a decay phenomenology around the divide
of BSMs with and without Supersymmetry (SUSY). In fact, among the possible BSM theories, SUSY remains one of the favourite ones. However, while its minimal realisation, the Minimal Supersymmetric Standard Model (MSSM), has been under close experimental scrutiny lately, through direct searches for both its sparticle and Higgs states,
much less effort has gone into testing non-minimal SUSY scenarios. Amongst the latter, a particular role is played by the Next-to-MSSM (NMSSM). Further, a slight variation of the latter, known as the New Minimal Supersymmetric Standard Model (nMSSM), has recently also undergone significant phenomenological scrutiny. All such SUSY scenarios are built upon a Higgs sector which is
essentially one particular realisation of a 2-Higgs Doublet Model (2HDM), with (NMSSM, nMSSM) or without (MSSM) an additional
Higgs singlet field. Thus, if one abandons the paradigm of SUSY, it is natural the examine generic 2HDMs. In fact,
all such extended Higgs models are capable of producing the Higgs cascade decays which are of  interest here, apart from the MSSM, which we will then not test. Regarding the others, we will tackle them in turn.

This paper is thus organised as follows.   In the next section, we shall review the discussed theoretical models (in separate subsections)
while in the following one we will describe the experimental analyses exploiting the mentioned signatures. Our results, obtained by
confronting predictions from the former with constraints from the latter, are presented in Section~\ref{sec:results}.
Finally, we conclude in Section~\ref{sec:conclusion}.

\section{Models}
\label{sec:models}
We now briefly review a number of models, all based around the Two Higgs Doublet Model (2HDM).
Whilst they differ in their input parameters and number of fields, they all share the ability to produce small mass (pseudo)scalars, with sizeable Higgs-to-Higgs couplings.

A scan over parameter space was performed for each model, targetting scenarios with low mass \aone\ or \hone.
Scans were subjected to many existing experimental constraints.
SM Higgs searches and measurements can be used to place indirect limits on our models.
All scans used \hb\ 4.3.1~\cite{Bechtle:2008jh,Bechtle:2011sb,Bechtle:2013gu,Bechtle:2013wla,Bechtle:2015pma} to implement current Higgs exclusion limits.
\hs\ 1.4.0~\cite{Bechtle:2013xfa} was used to apply measured Higgs signal rate constraints in a variety of channels.
This was run in peak-centered mode, with a Gaussian probability distribution, and requiring the overall $p > 0.05$.
For the NMSSM and nMSSM scans, we also consider the individual requirements on $ZZ/\diphoton/\bbbar$ Higgs signal rates which can result in different exclusion regions.
This will be discussed further in Section~\ref{sec:results}.

We also consider non-Higgs constraints, including those on flavour variables, the anomalous muon magnetic moment \amm, and dark matter (DM) relic density \dmrelic.
All scans use \micromegas~\cite{micromegas} to implement the latter constraint.
We apply a ``relaxed'' set of constraints, requiring points to pass all constraints but allowing any $\Delamu > 0$ and $\dmrelic < 0.131$, and ignoring constraints on \rd, \rds~\cite{HFAG}.
This allows for future developments and changes in those calculations, whilst still accommodating some BSM contribution.
This does not significantly modify the results in Section~\ref{sec:results}.

\subsection{Type I and II Two Higgs Doublet Model (2HDM)}
\label{subsec:2HDM}

The Two Higgs Doublet Model (2HDM) represents one of the most economical extension of the SM Higgs sector, providing a simple, yet comprehensive, framework for studying extended patterns of EW symmetry breaking.
In the 2HDM a second complex Higgs doublet with the same quantum numbers of the SM one is added to the SM Higgs sector.
The scalar spectrum of the 2HDM is thus enlarged to include two CP even states, denoted as $h$ and $H$ (with $m_h<m_H$), a CP odd state, $A$, and a charged Higgs, $H^\pm$. In general, the role of the SM like Higgs boson can be played by either $h$ or $H$.

Denoting the two Higgs doublets as $\Phi_{1,2}$, the most generic scalar potential of the 2HDM that respects a $\mathbb{Z}_2$ symmetry distinguishing $\Phi_1$ and $\Phi_2$ can be expressed as~\cite{Davidson:2005cw}

\begin{equation}
\label{eq:2hdm_pot}
\begin{split}
  V_{\mathrm{2HDM}}& = \sum_{i=1}^2 m_{ii}^2\Phi_i^\dag \Phi_i - [m_{12}^2 \Phi_1^\dag \Phi_2 + h.c.]+\sum_{i=1}^2 \lambda_i (\Phi_i^\dag \Phi_i)^2 + \\
 & + \lambda_3 (\Phi_1^\dag \Phi_1)(\Phi_2^\dag \Phi_2) + \lambda_4 (\Phi_1^\dag \Phi_2)(\Phi_2^\dag \Phi_1)  + \left[\frac{1}{2}\lambda_5 (\Phi_1^\dag \Phi_2)^2 +h.c.\right].\\
 \end{split}
\end{equation}

The imposition of a $\mathbb{Z}_2$ symmetry, together with the assignment to the right handed SM quarks of a defined $\mathbb{Z}_2$ quantum number, is necessary so as to avoid Higgs mediated flavour changing neutral currents (FCNC).
Note that the $\mathbb{Z}_2$ breaking term $m_{12}^2$ is generally tolerated, since it breaks the $\mathbb{Z}_2$ symmetry softly, {\emph i.e.} the symmetry is restored in the UV~\cite{Ginzburg:2004vp}.

In the potential of eq.~(\ref{eq:2hdm_pot}) the parameters $\lambda_{1-4}$, $m_{11}^2$ and $m_{22}^2$ are real numbers, while $m_{12}$ and $\lambda_5$ are in principle allowed to be complex valued numbers. However, complex parameters that cannot be made real through a suitable transformation give rise to CP violation in the Higgs sector. Since we are not interesting in the study of these effects, in the following we will consider all the parameters of the potential to be real numbers.

Starting from the scalar potential of eq.~(\ref{eq:2hdm_pot}), various 2HDM realisations can then be formulated according on how the SM fermions couple to the two Higgs doublets. In particular we will focus in our analysis on the so called Type I and Type II 2HDMs.
In Type I 2HDM all the SM fermions, up and down type quarks and down type leptons, couple to only one doublet while in Type II down type quarks and leptons couple to one doublet and up type quarks to the other doublet.

In order to scan the 2HDM parameter space we have used the package \verb#2HMDC#~\cite{Eriksson:2009ws} with input parameters defined in the mass basis.
In this basis the free model parameters are the physical masses of the four scalar states ($m_h$, $m_H$, $m_A$, $m_{H^\pm}$), the ratio of the two doublets vacuum expectation values ($\tan\beta=v_2/v_2$), $m_{12}^2$, and $\sin(\beta-\alpha)$, with $\alpha$ the mixing angle between the two scalar states. The parameter ranges used for the scan are indicated in Table~\ref{tab:2HDMphenoParamRange}.
The \verb#2HMDC# package imposes basic theoretical constraints, such as stability of the potential, tree level unitarity, and consistency with the S, T, and U EW parameters.
Finally \superiso~\cite{superiso} was used to check compatibility with current flavour constraints.
However failing points were not explicitly excluded to increase the overall scan efficiency.

\begin{table}[h!]
    \centering
    \begin{tabular}{cl}
        \multicolumn{2}{c}{$h = \hdisc$}\\
        \hline
        Parameter & Range \\
        \hline
        $m_{h}$ & 124 -- 128 GeV \\
        $m_{H}$ & 128 -- 1000 GeV \\
        $m_{A}$ & 3.5 -- 40 GeV \\
        $m_{H^\pm}$ & 128 -- 1000 GeV \\
        $\tan\beta$ & 0.5 -- 50 \\
        $m_{12}^2$ & 10 -- $10^5$  GeV$^2$\\
        $|\sin(\beta-\alpha)|$ & 0.9 -- 1 \\
        \hline
    \end{tabular}
    \quad\quad\quad
        \begin{tabular}{cl}
        \multicolumn{2}{c}{$H = \hdisc$}\\
        \hline
        Parameter & Range \\
        \hline
        $m_{h}$ & 3.5 -- 124 GeV \\
        $m_{H}$ & 124 -- 128 GeV \\
        $m_{A}$ & 3.5 -- 40 GeV \\
        $m_{H^\pm}$ & 128 -- 1000 GeV \\
        $\tan\beta$ & 0.5 -- 50 \\
        $m_{12}^2$ & 10 -- $10^5$ GeV$^2$ \\
        $|\cos(\beta-\alpha)|$ & 0.9 -- 1 \\
        \hline
    \end{tabular}
    \caption{2HDM parameters and their ranges used for the scans. Left table for $m_h=125$ GeV, right for $m_H=125$ GeV.}
    \label{tab:2HDMphenoParamRange}
\end{table}

\subsection{Next-to-Minimal Supersymmetric Standard Model (NMSSM)}
\label{subsec:nmssm}
The Next-to-Minimal Supersymmetric Standard Model (NMSSM)~\cite{Ellwanger:2009dp} is a simple extension of the MSSM, which adds a singlet $S$ to its superpotential.
Originally proposed to solve the $\mu$-problem of the MSSM, the NMSSM has gained renewed interest as additional tree-level contributions to the Higgs mass alleviates the need for large loop contributions to achieve its measured value, thus possibly allowing a more natural sparticle spectrum~\cite{Ellwanger:2011aa,Jeong:2012ma,Cao:2012fz,Agashe:2012zq,Barbieri:2013hxa,Badziak:2013bda,Barbieri:2013nka}.

The inclusion of a new singlet scalar naturally also leads to more physical scalar particles: one scalar and one pseudoscalar will be added giving in total three scalars ($h_{1,2,3}$), two pseudoscalars ($a_{1,2}$), and the usual charged Higgs \hc.
A novel feature is that the discovered Higgs can be assigned to either \hone\ or \htwo.
The latter possibility was found to be excluded in the MSSM by~\cite{Arbey:2012dq,Arbey:2012bp} due to a combination of flavour observables and LHC searches for scalars decaying to $\tau\tau$ pairs, though one might add that more recently~\cite{Bechtle:2016kui} claims there still is a very constrained possibility that the heavier scalar is the discovered one in the phenomenological MSSM.

The inclusion of the extra singlet superfield results in a modified superpotential,
\begin{equation}\label{eq:SuperPot}
  W_{\mathrm{NMSSM}} \supset \lam \widehat{S}\widehat{H}_u\widehat{H}_d + \frac{\kap}{3}\widehat S^3,
\end{equation}
where \lam\ and \kap\ are dimensionless coupling constants, and we have assumed a $\mathbb{Z}_3$ invariant model.
The rest of the superpotential is formed from the usual Yukawa terms for quarks and leptons as in the MSSM.
Further, one needs to add the corresponding soft supersymmetry breaking terms in the scalar potential,
\begin{equation}\label{eq:SoftHiggs}
\begin{split}
V^\mathrm{NMSSM}_{\mathrm{soft}} &\supset m_{S}^2 | S |^2 + \left( \lambda A_\lambda H_u H_d S + \frac{\kap}{3} A_{\kappa}S^3 +  \mathrm{h.c.}\right)\; , \\
\end{split}
\end{equation}
where $m_S$, \alam\ and \akap\ are dimensionful mass and trilinear parameters, and one also has the other usual MSSM soft SUSY breaking terms.

\begin{figure}[!htbp]
    \centering
    \includegraphics[width=\textwidth]{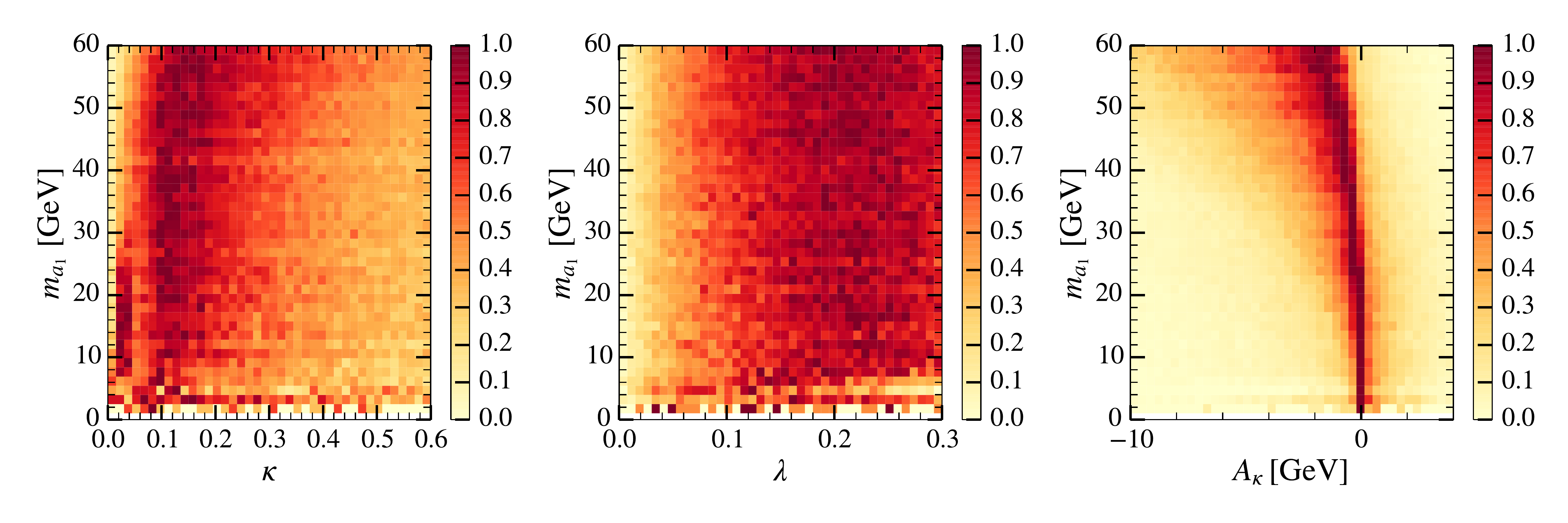}
    \caption{Heatmaps of \maone\ as a function of several NMSSM input parameters: \kap, \lam, and \akap. Each horizontal bin is normalised such that the largest bin in each row has contents = 1. Relaxed constraints have been applied, apart from those on Higgs signal rates.}
    \label{fig:pheonMAoneInput}
\end{figure}

As the masses of the singlet dominated scalar and pseudoscalar are essentially free parameters, it opens the possibility for them to be very light.
If the singlet component of \aone\ is large enough, then such light particles can easily escape all exclusion limits from earlier searches.
We briefly consider \maone\ as a function of selected input parameters, showing  the results in Fig.~\ref{fig:pheonMAoneInput}.
Scan details are explained below.
Relaxed constraints have been applied, apart from those on Higgs signal rates.
Each horizontal bin is normalised such that the largest bin in each row has contents = 1.
This allows one to see which value(s) of input parameter are preferred for a given \maone.
There are a few salient features to note.
Most strikingly, panel (a) shows that $\akap \sim 0$ or slightly negative is highly favoured for a light \aone\ scenario.
Panel (b) indicates some preference for $\kap \lesssim 0.3$, with another ``hotspot'' of points at $\kap \sim 0.02 - 0.04$.
Panel (c) also shows a weak preference for a fairly small $\lam \sim 0.15$.

Whilst a scalar with mass $\sim 125\ \GeV$ is easily achievable in the NMSSM, it is useful to momentarily review its dependence on the model input parameters.
A scalar with mass $125 \pm 3\ \GeV$ is achievable over the parameter range scanned.
\begin{figure}[!htbp]
    \centering
    \includegraphics[width=\textwidth]{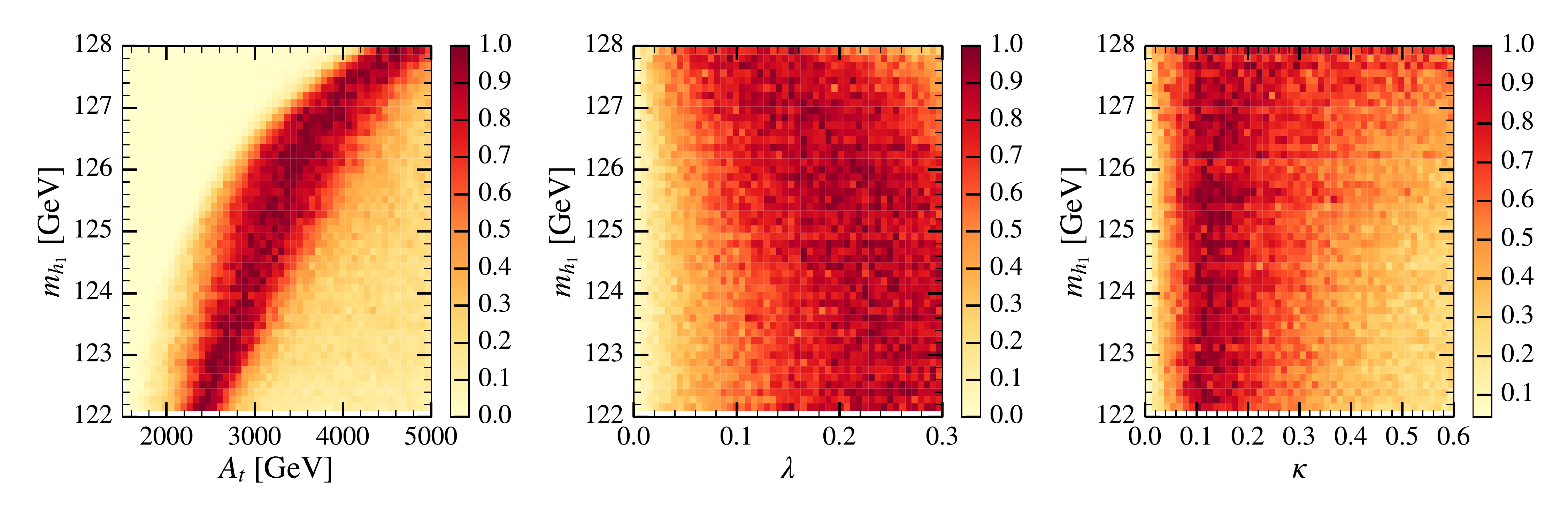}
    \caption{Heatmaps of \mhone\ as a function of several NMSSM input parameters: $A_t$, $\lambda$, and $\kappa$. Each horizontal bin is normalised such that the largest bin in each row has contents = 1. Relaxed constraints have been applied, apart from those on Higgs signal rates.}
    \label{fig:phenoMHoneInput}
\end{figure}
Fig.~\ref{fig:phenoMHoneInput} shows the dependence of \mhone\ on selected parameters where there are noticeable trends.
Relaxed constraints have been applied, apart from those on Higgs signal rates.
In particular, $A_t$ (left panel) sets an upper limit on \mhone\ through its effect on the stop mixing which in turn effects the loop contributions to the Higgs mass.
Additionally, smaller values of \lam\ (central panel) tend to push \mhone\ to larger values. It may seem surprising that smaller \lam\ allows larger \mhone, while the NMSSM specific contribution to \mhone\ is proportional to \lam. But in our case all large \lam\ are already excluded by the signal rate constraints and \mhone\ only shows a clear growth with \lam\ for \lam$>$0.4, below that one also has to remember that \lam\  affects the mixing of the scalars and thus can have a more complicated impact on \mhone. We also see that smaller values of \kap$\sim 0.1$ -- $0.3$ (right panel)  are preferred in order to satisfy signal rate constraints for \hone.

There have been numerous studies of light pseudoscalars in the NMSSM and their discovery prospects, see, e.g., \cite{Accomando:2006ga,Almarashi:2010jm,Almarashi:2011bf,Ellwanger:2005uu,Ellwanger:2003jt,Ellwanger:2001iw,Hugonie:2001ib,Belyaev:2008gj,Forshaw:2007ra,Belyaev:2010ka,Moretti:2006sv,Moretti:2006hq,Almarashi:2011hj,Ellwanger:1999ji,Djouadi:2008uw,Mahmoudi:2010xp,Cao:2013gba,Bomark:2014gya,Bomark:2015fga}
 but the present study is the first attempt to investigate the impact on the NMSSM parameter space from LHC searches for light pseudoscalars.
For our analysis, we have performed scans for both the $\mathbb{Z}_3$-invariant NMSSM (hereafter referred to as just the NMSSM), and a GUT inspired NMSSM.
In the latter, one has a common parameter for all scalar masses ($m_0$), a common parameter for all trilinear parameters except \alam\ and \akap\ ($A_0$), and a typical GUT relation between the gaugino masses ($M_2= M_1/2 = M_3/3=\mhalf$ at the EW scale).
The singlet pseudoscalar mass parameter,  $M_p$, is used as an input parameter in the GUT scan instead of \akap, requiring input parameters to be specified at the EW scale to be effective.
The parameter ranges for the NMSSM scan are described in Table~\ref{tab:phenoParamRange}, while the ranges in the GUT inspired scan are given in Table~\ref{tab:GUTParamRange}; here two scans were made, one (reduced range) focusing on the region with large \lam\ and small \tanbeta\ to optimise the NMSSM specific contribution to the Higgs mass, and one broader (extended range) to ensure no possibility was missed.
\begin{table}[!htbp]
    \centering
    \begin{tabular}{cl}
        \hline
        Parameter & Range \\
        \hline
        $\lambda$ & 0 -- 0.3 \\
        $\kappa$ & 0 -- 0.6 \\
        $\tan\beta$ & 10 -- 30 \\
        $\mu_{\rm{eff}}$ & 180 -- 220 \GeV \\
        $A_{\lambda}$ & 100 -- 4000 \GeV \\
        $A_{\kappa}$ & -10 -- 4 \GeV \\
        $A_t$ & 1500 -- 5000 \GeV \\
        $A_b$ & 500 -- 2500 \GeV \\
        \hline
    \end{tabular}
    \begin{tabular}{cl}
        \hline
        Parameter & Range \\
        \hline
        $M_1$ & 150 \GeV \\
        $M_2$ & 300 \GeV \\
        $M_3$ & 250 -- 2500 \GeV \\
        $M_{U_1} = M_{U_2} = M_{U_3}$ & 500 -- 2500 \GeV \\
        $M_{D_1} = M_{D_2} = M_{D_3}$ & 500 -- 2500 \GeV \\
        $M_{Q_1} = M_{Q_2} = M_{Q_3}$ & 800 -- 2500 \GeV \\
        $M_{E_{1/2/3}} = M_{L_{1/2/3}}$ & 1000 \GeV \\
        $A_{e/\mu/\tau}$ & 2500 \GeV \\
        \hline
    \end{tabular}
    \caption{NMSSM parameters and their ranges used for the scans. All parameters are specified at the SUSY scale.}
    \label{tab:phenoParamRange}
\end{table}

\begin{table}[!htbp]
    \centering
    \begin{tabular}{cll}
            \hline
         Parameter & Extended range & Reduced range  \\
    \hline
    $m_0$ (GeV) & 200 -- 2000 & 200 -- 2000 \\
    \mhalf\,(GeV)  & 100 -- 2000 & 100 -- 1000 \\
    $A_0$ (GeV)  &  $-5000$ -- 5000 & $-3000$ -- 3000 \\
    $\mu_{\rm eff}$ (GeV) & 50 -- 1000 & 100 -- 200 \\
    \tanbeta & 1 -- 30 & 1 -- 6 \\
    \lam  & 0.01 -- 0.7 & 0.4 -- 0.7  \\
    \kap  & 0.01 -- 0.7 & 0.01 -- 0.7 \\
    \alam\,(GeV) & 200 -- 2000 & 200 -- 1000 \\
    $M_p$ (GeV) & 3 -- 140 & 3 -- 140 \\
    \hline
    \end{tabular}
    \caption{Parameter ranges used in the GUT inspired NMSSM scans.  All parameters are specified at the EWK scale.}
    \label{tab:GUTParamRange}
\end{table}
All the NMSSM scans use \nt\ (v4.9.3 for the NMSSM, v4.6.0 for the GUT inspired scan)~\cite{Ellwanger:2004xm,Ellwanger:2005dv,Belanger:2005kh} to calculate sparticle spectra and ensure consistency with LEP and LHC exclusions.
The GUT inspired scan also uses \texttt{MultiNest}-v2.18~\cite{Feroz:2008xx}, and \texttt{SuperIso}-v3.3 to check constraints from B physics.
\nt\ includes both Higgs exclusion and signal strength constraints from experimental results, based on \lilith~\cite{Bernon:2015hsa} database version 15.09.
Flavour constraints have also been implemented in \nt~\cite{Domingo:2015wyn}, and points are checked against these constraints.

In order to use \hs\ with the output from \nt, we add a \verb=DMASS= block to the SLHA file to represent theoretical uncertainties on the \hdisc\ mass.
This is set to 2 \GeV\ for both \hone\ and \htwo.
Additionally, \hs\ was modified to ensure that either \hone\ or \htwo\ was correctly assigned to \hdisc\ by increasing \verb=assignmentrange_massobs= to 2.0 in \verb=usefulbits_HS.f90=.

\subsection{New Minimal Supersymmetric Standard Model (nMSSM)}
\label{subsec:littleNmssm}

In the previous section we have described the properties of the $\mathbb{Z}_3$ invariant NMSSM. However, a general 2HDM+S superpotential might not posses this accidental symmetry. A different realisation, called the new Minimal Supersymmetric Standard Model (nMSSM), possesses instead a discrete R-symmetry that forbids a cubic singlet term in the superpotential but allows for tadpole terms. While the field content of the nMSSM is the same as that of the $\mathbb{Z}_3$ invariant NMSSM, the phenomenology can be quite different due to the different superpotential and soft SUSY breaking terms.

The first striking feature of the nMSSM is the absence of a mass term for the pure singlino, whose mass can be raised up to $\sim 75$ GeV only via mixing effects. The singlino is thus naturally light and the LSP, which generally contains a large singlino component, can have a mass lighter than $\sim 5$ GeV, leading to a quite different phenomenology for the nMSSM in both collider and DM searches.

The Higgs sector of the nMSSM superpotential reads~\cite{Ellwanger:2009dp} (in contrast to eq.~\ref{eq:SuperPot})
\begin{equation}
W_{\rm nMSSM} \supset \lambda \widehat{S}\widehat{H}_u\widehat{H}_d + \xi_F\widehat{S}\;,
\end{equation}
to which the usual Yukawa terms are added.
The corresponding soft SUSY breaking terms are very similar to eq.~\ref{eq:SoftHiggs}, but removing the $\frac{\kap}{3} A_{\kappa}S^3$ term and introducing a tadpole term:
\begin{equation}
\begin{split}
V^\mathrm{nMSSM}_{\mathrm{soft}} &\supset m_{S}^2 | S |^2 + \left( \lambda A_\lambda H_u H_d S + \xi_S S +  \mathrm{h.c.}\right)\; ,  \\
\end{split}
\end{equation}
where $\xi_F$ and $\xi_S$ are $\mathcal{O}(M_{SUSY}^2)$ and $\mathcal{O}(M_{SUSY}^3)$ terms which avoid domains walls and stability problems of the nMSSM (see~\cite{Ellwanger:2009dp}).

Our reinterpretation of the constraints arising from low mass 8 TeV scalar searches will be based on the results presented in a recent paper~\cite{Barducci:2015zna} that reviews the status of the nMSSM after the first run of the LHC and highlights the prospects for this model for the 13 TeV run of the CERN machine.
Referring to~\cite{Barducci:2015zna} for more details, we summarise here the major details of the parameter scan and of the constraints imposed.
\nt\ has been used to scan over the following parameters:
\begin{equation}
m_0, \; M_{1/2}, \; A_0, \; \mu, \; \tan_\beta, \; \lambda, \; \xi_F, \; \xi_S, \; A_\lambda
\end{equation}
all defined at the GUT scale except $\tanbeta$, defined at $M_Z$, and $\lambda$ and $\mu$, both defined at the SUSY scale.
We impose the following universal soft terms conditions at the GUT scale:
\begin{equation}
\left\{\begin{array}{l}
m_Q = m_U = m_D = m_L = m_E \equiv m_0 \\
A_u = A_d = A_e \equiv A_0 \\
M_1 = M_2 = M_3 \equiv M_{1/2} \;.
\end{array}\right.
\end{equation}

Regions of the parameter space where sparticles are out of the LHC reach have been discarded, thus only focusing on regions with interesting prospects at present and future colliders.
Constraints on direct sparticle searches at LEP, Tevatron, and the LHC have been implemented via the \texttt{SModelS}~\cite{Kraml:2014sna,Kraml:2013mwa} and \texttt{MadAnalysis5}~\cite{Conte:2012fm,Conte:2014zja,Dumont:2014tja} packages.

In~\cite{Barducci:2015zna}, three regions compatible with the aforementioned combination of theoretical, cosmological and collider constraints were identified.
In two of them the LSP has a mass of $\sim 45$ GeV and $\sim 70$ GeV respectively, while a third region features a light LSP, $m_{LSP}<5$ GeV.
This is the only region with a light spin 0 state, $a_1$, in the mass range of interest for this paper.
In particular one has $m_{a_1}\sim 2 m_{\tilde \chi^0_1}$, which ensures an efficient annihilation in the early Universe and thus provides a relic abundance compatible~\footnote{Regions where the DM relic abundance is below the experimental value have been considered as valid.} with the value measured by the Planck collaboration~\cite{Adam:2015rua}.
Within this region, there are two different subregions, denoted as 1A and 1B.
Region 1A is characterised by a small $m_0$ and $M_{1/2}$, both below 1 TeV, whilst region 1B has a small $M_{1/2}$ ($<500$ GeV) and large $m_0$ ($>$ 4 TeV).
Their full parameter ranges are reported in Table~\ref{tab:littlenmssmphenoParamRange}.

Unlike the NMSSM, in both these regions the role of the SM Higgs boson is played by $h_2$, with $h_1$ having a mass between 35 and 70 GeV.
As previously mentioned, $a_1$ is the lightest of the Higgs states which has a dominant singlino component, while the remaining heavier Higgs are decoupled.
In particular region 1B features an extremely light gluino, with $m_{\tilde{g}} \lesssim 1.2\ \TeV$, and is almost nearly excluded by run 1 searches. LHC results for stop and slepton searches also strongly constrains region 1A, via , which are light in this part of the parameter space where $m_0$ is small.

\begin{table}[h!]
    \centering
    \begin{tabular}{cl}
        \multicolumn{2}{c}{\textbf{Region 1A}}\\
        \hline
        Parameter & Range \\
        \hline
        $\tan\beta$ & 6.6 -- 10\\
        $\lambda$ & 0.33 -- 0.53\\
        $\mu$ & 240 -- 400 GeV \\
        $m_0$ & 0 -- 1080 GeV\\
        $M_{1/2}$ & 630 -- 1200 GeV\\
        $A_0$ & -1700 -- 50 GeV\\
        $A_\lambda$ & 1400 -- 6000 GeV\\
        $\xi_F$ & 10 -- 100 GeV$^2$\\
        $\xi_S$ & -6$\times 10^4$ -- 2$\times 10^4$ GeV$^3$\\
        \hline
    \end{tabular}
    \quad\quad\quad
        \begin{tabular}{cl}
        \multicolumn{2}{c}{\textbf{Region 1B}}\\
        \hline
        Parameter & Range \\
        \hline
        $\tan\beta$ & 6 -- 8\\
        $\lambda$ & 0.49 -- 0.52\\
        $\mu$ & 350 -- 430 GeV \\
        $m_0$ & 4040 -- 4800 GeV\\
        $M_{1/2}$ & 280 -- 440 GeV\\
        $A_0$ & 6700 -- 7900 GeV\\
        $A_\lambda$ & 7000 -- 7900 GeV\\
        $\xi_F$ & -1.5$\times 10^4$ -- -1.4$\times 10^4$ GeV$^2$\\
        $\xi_S$ & -1.9$\times 10^7$ -- -1.6$\times 10^7$ GeV$^3$\\
        \hline
    \end{tabular}
    \caption{nMSSM parameter ranges surviving the scan described in the text. Left table for region 1A, right for region 1B.}
    \label{tab:littlenmssmphenoParamRange}
\end{table}

\section{New Experimental Analyses}
\label{sec:experiments}

There are several recent experimental analyses searching for light bosons which may impinge on the parameter space of the aforementioned 2HDM and NMSSM/nMSSM scenarios.
We provide an overview of the ones most relevant to this investigation, categorised by their final state.
Note that while we refer to \aone, it should be understood that this can refer to a generic light boson, \aone\ or \hone.

For scenarios where $\maone << m_h$, a common theme is that of ``boosted'' topologies, where the \aone\ is significantly boosted, and therefore its decay products are highly collimated~\cite{Curtin:2013fra}.
The separation is of the order $\DR \sim 2\maone / p_T^a \sim 4\maone/m_h$, where we have assumed that each $a_1$ has a transverse momentum $p_T^a \sim m_h / 2$.
For $\maone \sim 8\ \GeV$, we therefore expect $\Delta R \sim 0.3$.
Analyses must therefore take care to ensure standard isolation criteria do not inadvertently quash any potential signal.
At larger \maone, the \aone\ is no longer highly boosted, and there is good separation between its decay products.
Standard reconstruction techniques can therefore be used.
The intermediate region, $\maone \sim 15$ -- $20\ \GeV$, proves the most challenging since the decay objects are neither neatly collimated, nor well separated.

\subsection{Adapting Experimental Limits}
\label{subsec:adaptLimit}

One can adapt the limit from a search for one final state to place a limit on another, given a relationship between the corresponding final states.
The channel widths are given in~\cite{Djouadi:2005gj}.
Since all leptons and down-type quarks couple to the same doublet in the models under consideration, there is no \tanbeta\ dependence and the conversion is simple.
For $\mu\mu \to \tau\tau$:

\begin{equation}
\frac{BR(\aone \to \tau\tau)}{BR(\aone \to \mu\mu)} = \frac{m^2_{\tau}\:\beta(\mtau, \maone)}{m^2_{\mu}\:\beta(\mmu, \maone)}
\end{equation}
where
\begin{equation}
\beta(m_X, \maone) = \sqrt{1 - \left(\frac{2m_X}{\maone}\right)^2}
\end{equation}
is the velocity factor.

For $\bbbar \to \tau\tau$:
\begin{equation}
\frac{BR(\aone \to \tau\tau)}{BR(\aone \to \bbbar)} = \frac{m^2_{\tau}\:\beta(\mtau, \maone)}{3\bar{m}^2_{b}\:\beta(\bar{m}_b, \maone) \times \left(1 + \Delta_{\qqbar} + \Delta_a^2\right)}
\end{equation}
where the radiative corrections are

\begin{align}
\Delta_{\qqbar} &= 5.67\frac{\alphasbar}{\pi} + (35.64 -1.35 N_f)\left(\frac{\alphasbar}{\pi}\right)^2 \\
\Delta_a^2 &= \left( \frac{\alphasbar}{\pi} \right)^2 \left(3.83 - \ln \frac{\maone^2}{m_t^2} + \frac{1}{6}\ln^2\frac{\mqbar^2}{\maone^2} \right)
\end{align}
where $N_f$ is the number of active light quarks; $\alphasbar$ is the running strong coupling constant; $\mqbar$ is the running quark mass in the $\overline{\mathrm{MS}}$ scheme; and $\alpha$ is the QED coupling constant.
The running parameters are evaluated at scale $\mu = \maone$ using~\cite{Agashe:2014kda,vanRitbergen:1997va,Botje:2010ay,Djouadi:2005gi,Curtin:2013fra}.

\subsection{\texorpdfstring{$4\tau$}{4tau}}
\label{subsec:experiments4tau}

For the mass region $2m_{\tau}$ -- $2m_b$, $BR(\aone \to \tau\tau)$ is expected to dominate in a Type II scenario with $\tanbeta \gtrsim 2$.
Ditau (or pairs of ditau) final states are therefore a natural search channel.
However due to the nature of the tau decay, it can be a difficult object to fully reconstruct in a boosted regime.
Taus can decay into 1, 3, or 5 charged particles (``prongs'') along with one or more neutral particles, including neutrinos.
The 1-prong and 3-prong decays modes make up $\sim 85\%$ and $\sim 15\%$, respectively, of all tau decays.
The multi-particle nature of the decay reduces the visible energy, making passing trigger thresholds and reconstruction more difficult then, \eg, $\aone \to \mu\mu$.

The CMS collaboration has published two analyses that search for $4\tau$ final states arising from pairs of low-mass boson decays~\cite{Khachatryan:2015nba,CMS-PAS-HIG-14-022}.
Whilst both look for $\hdisc \to 2\aone \to 4\tau$, and cover similar $\maone$ ranges, they utilise different analysis strategies to identify the boosted tau pairs.
Both analyses capitalise on the excellent muon reconstruction and low fake rates, and require two muons in an event.

The approach taken in~\cite{Khachatryan:2015nba} (CMS HIG-14-019) targets the tau 1-prong and muon decay modes.
Ditau pairs are selected by looking for a well-isolated muon with only one nearby track with $\PT > 2.5~\GeV$.
This forms a $\mu \mhyphen \mathrm{track}$ pair, and events are required to have 2 such pairs that are well separated.
Backgrounds are almost entirely from QCD heavy-flavour decays, since Drell-Yan, \ttbar, and diboson events are rejected by a same-sign requirements on the two muons.
The $\mu \mhyphen \mathrm{track}$ invariant mass, \mmutrack, is used as the discriminating variable.
A background template is formed from a QCD-rich sideband region, and fitted to the data along with signal template from MC to extract the size of any potential signal.
Upper limits on the total $\sigma \times BR$ range from 10.3 \pb\ at $\maone = 5\ \GeV$ down to 4.5 \pb\ at $\maone = 8\ \GeV$.

A complementary approach is taken in~\cite{CMS-PAS-HIG-14-022} (CMS HIG-14-022).
This analysis targets both the gluon fusion and WH production modes.
To target the boosted ditau pair, the standard tau reconstruction is modified.
The tau reconstruction is seeded by anti-\kT(with a 0.5 cone radius)~\cite{1126-6708-2008-04-063} jet candidates.
Candidate jets must have at least one muon constituent, which is removed before passing the remaining jet constituents to the tau reconstruction algorithm.
This tau must have $\PT>20\ \GeV$ and also pass isolation criteria.
Events are required to have at least one such muon-tau pair.
There is also an additional muon requirement, which must be well separated from the muon-tau pair.
This is designed to be sensitive to $W(\mu\nu)H$ production, or a muon from the other ditau pair in the gluon fusion and VBF production modes.
The analysis uses the $\mu \mhyphen \tau$ invariant mass to define a signal region, only considering events with $m_{\mu\mhyphen\tau} > 4\ \GeV$.
Upper limits on the total $\sigma \times BR$ range from $\sim 500\ \pb$ at $\maone = 5\ \GeV$ to 3.5 \pb\ at $\maone = 11\ \GeV$.

Both analyses are less powerful at smaller \maone\ as a consequence of using the effective ditau invariant mass as the discriminating variable.
Background events are characterised by small invariant mass, and thus there is a much larger overlap with a smaller \maone\ signal, thereby reducing its discriminating power.
In the case of HIG-14-022, the lack of any information below $4\ \GeV$ has a severe impact on the limit at small masses.
Additionally, the use of the visible ditau invariant mass means that there is no longer a clean, sharp peak on a continuous background, reducing the sensitivity of the searches compared to a fully reconstructible final state \eg \mumu.
\subsection{\texorpdfstring{\twotautwomu}{2tau2mu}}
\label{subsec:experiments2tau2mu}

This final state is a compromise between the large but less clean \tautau\ final state, and the much cleaner but rarer \mumu\ final state.
CMS and ATLAS have both published results looking for a \twotautwomu\ final state produced by light bosons~\cite{CMS-PAS-HIG-15-011,Aad:2015oqa}.
Both analyses look for resonances in the dimuon invariant mass distribution, and are triggered by an asymmetric dimuon requirement with similar \PT\ thresholds.
The CMS analysis targets a mass range $\maone = \left[20, 62.5\right]\ \GeV$, whilst the ATLAS result covers a range $\maone = [3.5, 50]\ \GeV$, optimising for $\maone = 5\ \GeV$.
The two analyses are therefore complementary.

Since the CMS analysis targets much larger values of \maone, the dimuon and ditau pairs will not be heavily boosted.
Therefore the standard hadronic tau reconstruction algorithm and isolation requirements can be used.
All four objects are required to be well separated, and events with additional isolated leptons or b-tagged jets are vetoed.
Requirements on the 4-body invariant mass and dimu-ditau mass difference are used to further enhance background rejection.
Both the reducible background (from jets faking leptons), and the irreducible background (from $ZZ \to 4\ell$), are modelled by Bernstein polynomials.
An upper limit on the \fourtau\ cross-section is set, ranging from $\sim 2\ \pb$ at $\maone \sim 20\ \GeV$ to $\sim 0.8\ \pb$ at $\maone \sim 60\ \GeV$.

In contrast, since the ATLAS analysis optimised for a much smaller mass, the kinematic and topological regime changes.
The dimuon and ditau pairs will now be heavily boosted, and akin to the CMS $4\tau$ analysis the ditau selection criteria avoids the use of a standard tau reconstruction algorithm, instead opting for a $\mu/e$ + tracks requirement.
The dimuon requirements include an isolation requirement, which is modified to remove the other muon.
This improves sensitivity at low \maone\ at the expense of reduced sensitivity at higher \maone.
Due to the mass range, the background estimation must now take into account various quarkonia resonances, as well as contributions from a continuum Drell-Yan background at smaller \maone, and \ttbar\ at large \maone.
The final upper limit on the \fourtau\ cross-section extends down to $< 1\ \pb$ for $\maone \sim 4\ \GeV$, but worsens at higher \maone, where it only reaches $\sim 20-30\ \pb$.
Since the selection criteria are not adapted for larger \maone, this is to be expected.

Interestingly, the ATLAS limit is better at smaller \maone\, despite the increase from the Drell-Yan background at smaller $m_{\mu\mu}$.
This is due to an increased signal efficiency. The lighter \aone\ receives a larger boost and therefore has a larger \PT\ on average, ensuring that more muons and tracks pass the trigger and selection requirements.
Whilst the same is also true in the \fourtau\ analyses, in those analyses the increase in signal efficiency is not sufficient to overcome the propinquity for background to lie at lower invariant masses.

\subsection{\texorpdfstring{$4\mu$}{4mu}}
\label{subsec:experiments4mu}

The region $\maone < 2\mmu$ sees a large increase in $BR(\aone\to\mumu)$.
Whilst not as large as $BR(\aone \to ss, gg)$, the dimuon final state is very clean with small systematic uncertainties.
Note that the other non-coloured final state, \diphoton, is still several orders of magnitude smaller than \mumu.
CMS has searched for a \fourmu\ final state~\cite{Khachatryan:2015wka}, targetting the pair production of very light (pseudo)scalars $\maone = [0.25, 3.55]\ \GeV$, each decaying to a pair of muons.
This analysis searches for two dimuon systems, with invariant masses compatible within detector resolution.
The muon pairing criteria takes into account situations in which the two muons are nearly parallel.
To reduce backgrounds from heavy-flavour decays, a modified muon isolation requirement is used, in which the other muon in the pair is excluded from the isolation sum.
The upper limits on the equivalent total \fourtau\ cross-section is $\sim 0.7$ -- $0.9\ \fb$.

\subsection{\texorpdfstring{$2b2\mu$}{2b2mu}}
\label{subsec:experiments2b2mu}
Focussing on higher masses, once the $2\mb$ threshold has been surpassed then this now becomes the dominant decay channel in the models under consideration (assuming $\tanbeta \gtrsim 2$ the for Type II models).
However a $4b$ search would have to overcome significant QCD backgrounds\footnote{Note however that ATLAS performed the first search for $4b$ in the $WH$ production channel at $\sqrt{s} = 13\ \TeV$~\cite{Aaboud:2016oyb}}.
Instead, requiring one \aone\ to decay to \mumu\ would allow one to use $m_{\mumu}$ as a powerful signal/background discriminant, improving search sensitivity.
CMS has performed a search for $h \to 2\aone \to 2b2\mu$ (HIG-14-041)~\cite{CMS-PAS-HIG-14-041}, covering a mass range $25 - 65\ \GeV$.
In this mass range the \aone\ is no longer boosted, and one can therefore utilise standard particle reconstructions algorithms.
This analysis required events to have two isolated muons, along with two b-tagged jets, with the 4-body invariant mass close to 125 GeV.
Signal and background functional templates are fit to the $m_{\mu\mu}$ distribution in data, where the background is dominantly $Z/\gamma$ + jets.
An upper limit is set, which is equivalent to a limit on the total \fourtau\ cross-section from 40 \fb\ to 100 \pb, assuming the relationships given in Section~\ref{subsec:adaptLimit}.
It should be noted that unlike other analyses, this limit is fairly constant with respect to \maone.

\section{Results}
\label{sec:results}
We now analyse how these new constraints affect the model parameter space by first considering the factors that influence the total cross-section, using the NMSSM as an example.
The total production cross-section predicted by a given model, $\sigma\times BR$, is decomposed as follows:

\begin{equation}
\label{eq:split}
\begin{split}
&\sigma \times BR (gg \to h \to 2\aone \to 2X 2Y) = \\
&\sigma^{8}_{SM}(ggh) \cdot g_{ggh}^2 \cdot BR(h \to 2\aone) \cdot BR(\aone \to 2X) \cdot BR(\aone \to 2Y) \cdot f
\end{split}
\end{equation}

where

\begin{itemize}
    \item $\sigma^{8}_{SM}(ggh)$ is the SM gluon-gluon fusion production cross-section at $\sqrt{s} = 8\ \TeV$ (19.27\ \pb ~for \mh\ = 125 \GeV~\cite{Heinemeyer:2013tqa})
    \item $g_{ggh}^2$ is the squared reduced $ggh$ coupling, with respect to the SM value (1 in the SM by definition)
    \item $BR(h \to 2\aone)$ is the branching ratio of $h$ to $2 \aone$
    \item $BR(\aone \to 2X)$ is the branching ratio of $\aone$ to $2 X$ where $X = \tau, \mu, \cdots$
    \item $f$ is a combinatorics factor: 1 if the final states $X$ and $Y$ are identical, 2 otherwise.
\end{itemize}

Note that we only consider gluon-gluon fusion production, since it is the dominant production mechanism.
There are several scenarios that involves light boson pair-production that we must consider: if $\hone = \hdisc$, then we could have $\hone /\htwo\to 2\aone$; if $\htwo = \hdisc$ then we could have $\htwo \to 2\aone/\hone$ or $\hone \to 2\aone$.

We now consider the squared reduced gluon-gluon-Higgs coupling, $g_{ggh}^2$, which \textit{a priori} is not constrained by the model.
Instead, it is heavily constrained by current experimental results.
If $h_i$ is assigned to be \hdisc, then Higgs coupling measurements mean it must be SM-like, \ie $g_{ggh_i}^2 \sim 1$.
If however it is not \hdisc, then current exclusion limits mean its production must be suppressed, \ie have a small $g_{ggh_i}^2$.
\begin{figure}[!htbp]
    \centering
    \includegraphics[width=\textwidth]{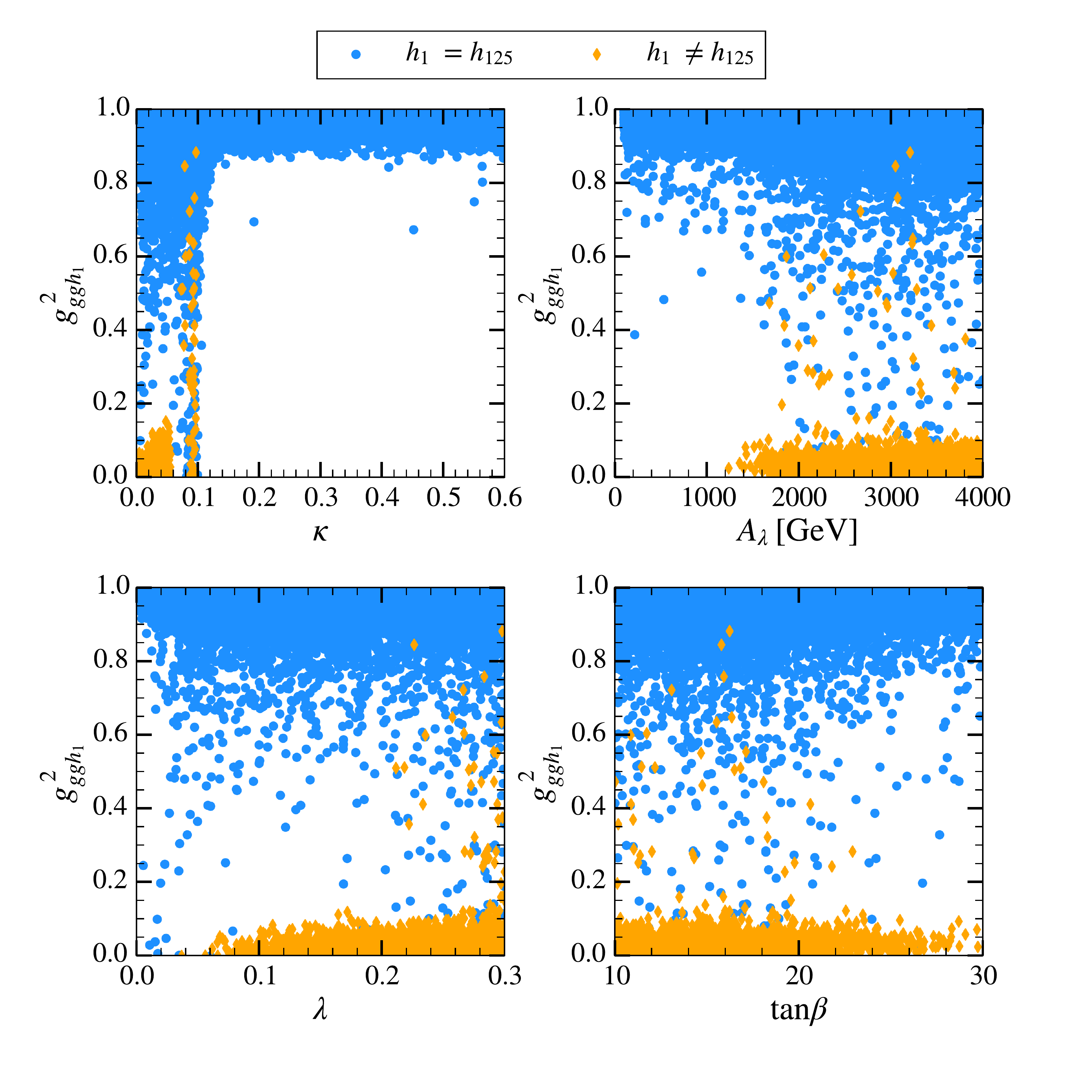}
    \caption{Squared $gg\hone$ coupling, $g_{gg\hone}^2$, normalised to the SM value, as a function of several input parameters in the NMSSM, for the cases when $\hone = \hdisc$ (blue circles) and when it is not the \hdisc\ (\ie when $\htwo = \hdisc$) (orange diamonds).}
    \label{fig:phenoH1ggrc2}
\end{figure}
The $gg\hone$ squared reduced coupling $g_{ggh_1}^2$ is shown in Fig.~\ref{fig:phenoH1ggrc2} as a function of several input parameters.
Blue points indicate models where $\hone = \hdisc$, whilst orange diamonds are models where $\htwo = \hdisc$.
Relaxed constraints have been applied, along with those on \hs\ and \hb.
We note that $g_{gg\hone}^2$ is far larger in models where $\hone = \hdisc$ compared with models where $\htwo = \hdisc$.
Additionally, in the former scenarios $g_{gg\hone}^2$ is easily able to reach 1 across the whole range of parameters scanned, in the latter it is confined to certain region of parameter space: particularly small \kap, and large \alam, with moderately sized \lam.
Generally, it is somewhat favoured to have $\hone = \hdisc$.
$g_{gg\htwo}^2$ follows a similar pattern: when $\htwo = \hdisc$ the reduced coupling can reach 1, whilst it is much smaller when $\hone = \hdisc$.
However, the former scenario is now confined to those aforementioned regions of parameter space: small \kap, and large \alam, with moderately sized \lam.

\begin{figure}[!htbp]
    \centering
    \includegraphics[width=\textwidth]{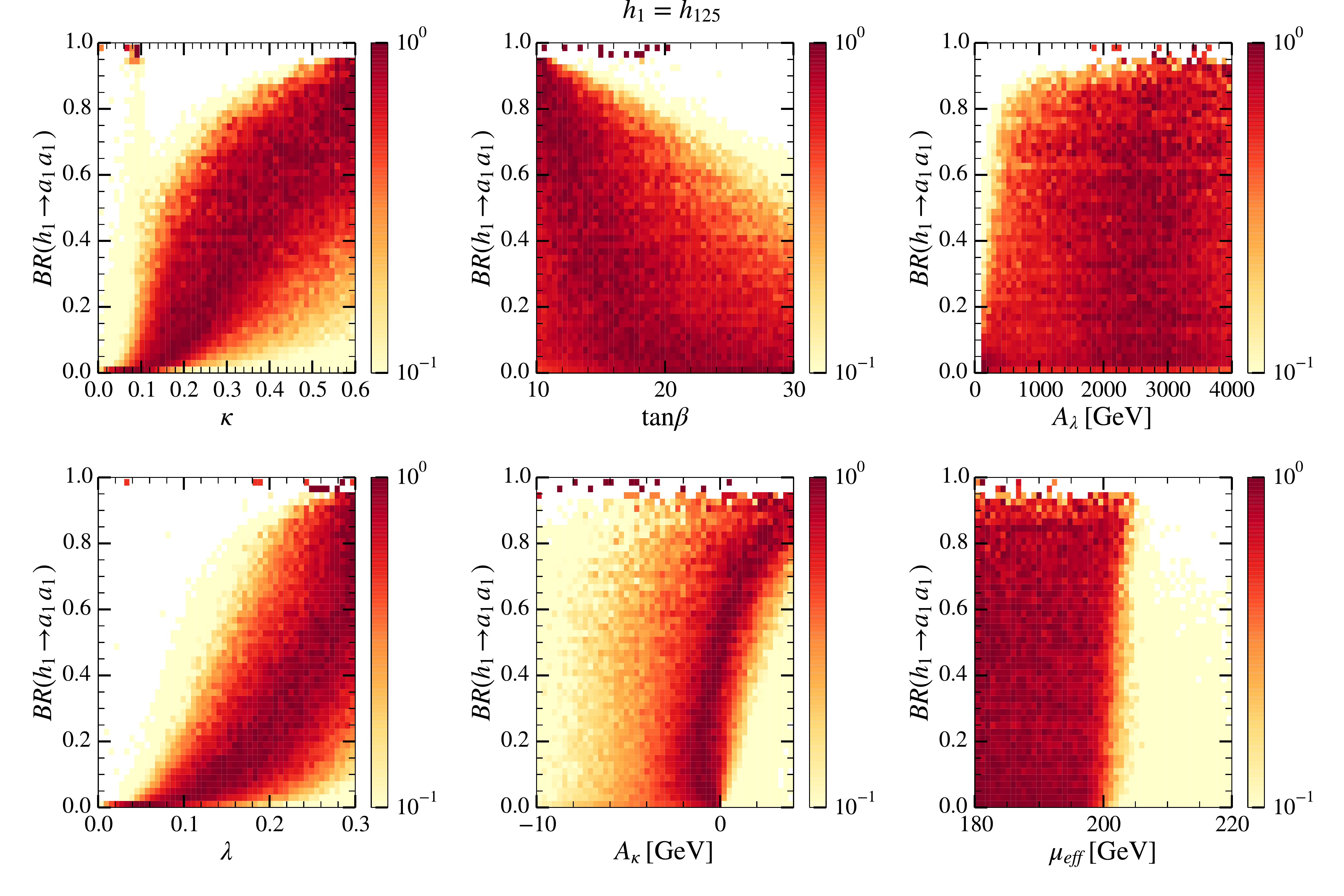}
    \caption{Heatmaps of $BR(\hone \to \aone\aone)$ for several model input parameters in the NMSSM, where $\hone = \hdisc$ and $\maone < 60\ \GeV$. All points pass all constraints except Higgs rate constraints from \hs\ or \nt, and we only require $\Delamu > 0$ and $\dmrelic < 0.131$. Each horizontal bin has been normalised such that the largest bin in each row has contents 1.}
    \label{fig:phenoBrh1a1a1Hist2D}
\end{figure}
The Higgs-to-Higgs branching ratio can also take on a range of values, and is again only limited by current Higgs measurements.
Fig.~\ref{fig:phenoBrh1a1a1Hist2D} shows heatmaps of $BR(\hone \to \aone\aone)$ against several model input parameters for points where $\hone = \hdisc$ and $\maone < 60\ \GeV$.
No Higgs coupling constraints have been applied from either \hs\ or \nt, but all other constraints have been applied.
Each plot is normalised such that each horizontal bin is scaled so that the largest bin in each row has contents 1.
This allows us to determine the sensitivity of a given $BR$ value against a model parameter.
Without any Higgs signal constraints, the $BR$ can take on any value.
We can also see clear features that show significant dependence of $BR(\hone \to \aone\aone)$ on these parameters, particularly \kap, \lam, and \akap; but also  some slight dependence on \tanbeta. The dependence on \kap\ and \lam\ can be understood due to the presence of $\lam^2$ and \kap\lam\ terms in the relevant coupling. Also \akap\ appears in that coupling, while the effect of \tanbeta\ is more indirect as it changes the relative importance of the $\lam^2$ and \kap\lam\ terms.

Adding in current Higgs coupling constraints requires a SM-like scenario for the SM decay channels and therefore a small $BR(\hdisc \to \mathrm{BSM})$, with the most recent combined fits from CMS and ATLAS constraining $BR(\hdisc \to \mathrm{BSM}) < 0.34$ at 2$\sigma$~\cite{Khachatryan:2016vau}.
A small $BR$ therefore primarily relies on a small $\kap \lesssim 0.3$ -- $0.4$, a small $\lam\ \lesssim 0.2$ -- $0.3$, and a negligible or slightly negative \akap. There is also a preference for large $\tanbeta \sim 10$ -- $25$, and large $\alam \sim 3\ \TeV$.
Note that we have not considered $\hdisc \to Z \aone$ decays, since their $BR$ are typically $\lesssim 10^{-8}$.

\begin{figure}[!htbp]
    \centering
    \begin{subfigure}[t]{0.48\textwidth}
        \includegraphics[width=\textwidth]{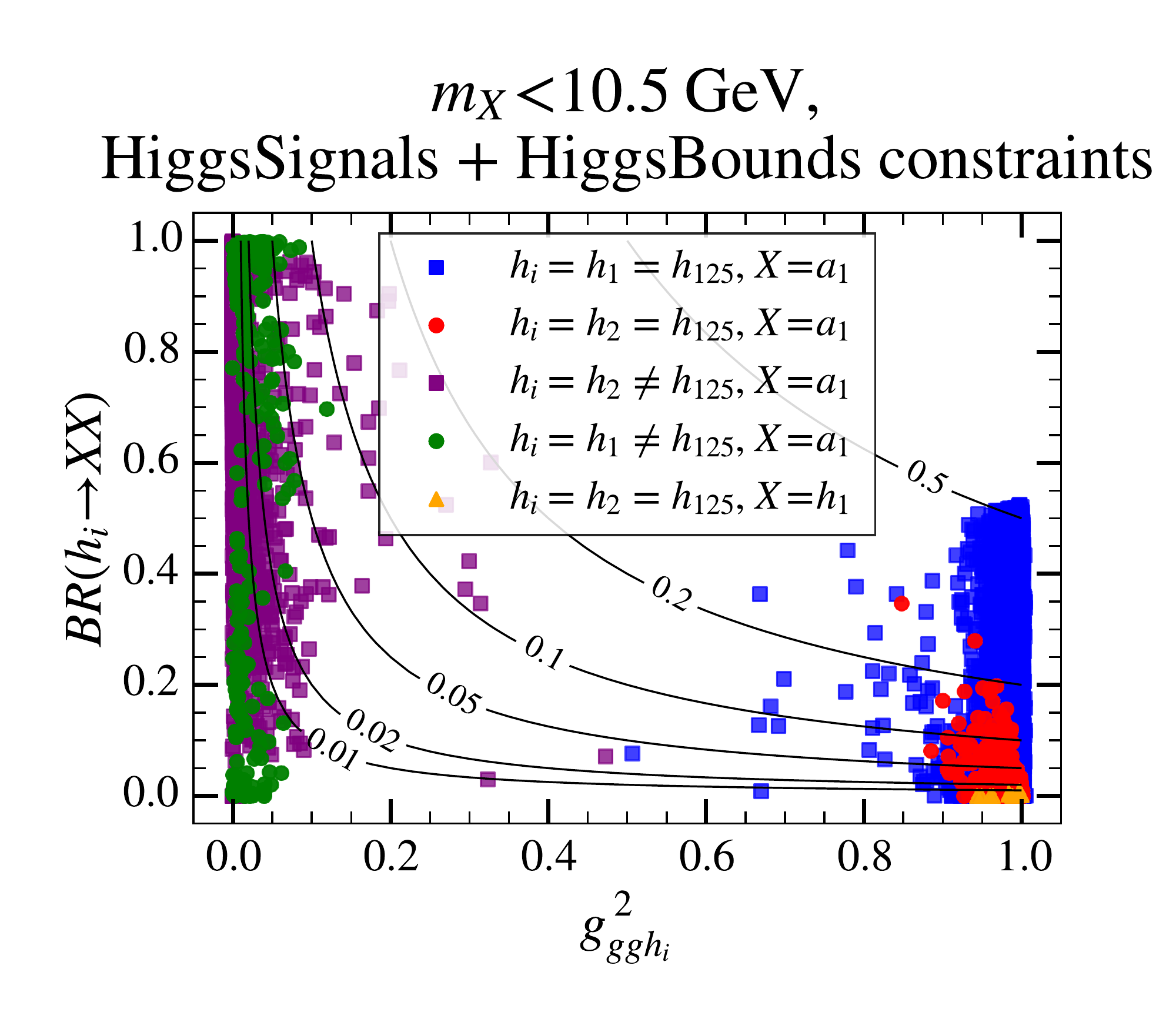}
        \caption{}
        \label{fig:phenoRC2BRHSHB}
    \end{subfigure}
    ~
    \begin{subfigure}[t]{0.48\textwidth}
        \includegraphics[width=\textwidth]{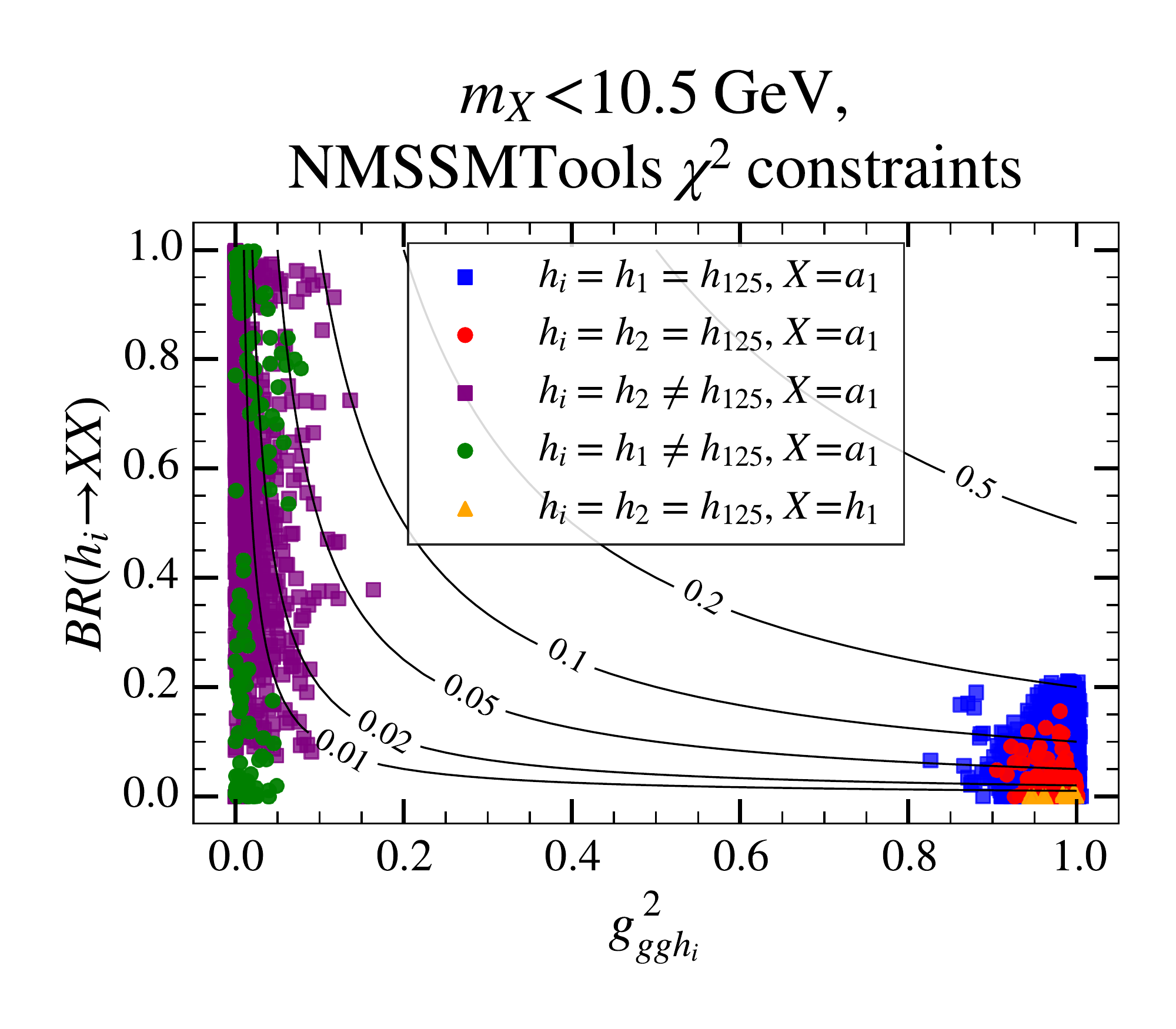}
        \caption{}
        \label{fig:phenoRC2BRNT}
    \end{subfigure}
    \caption{Scatter plots of Higgs-to-Higgs $BR$ against squared gluon-Higgs reduced coupling $g_{ggh}^2$ in the NMSSM, for different assignments of \hdisc\ and \aone. Contours of constant $BR \times g_{ggh}^2$ are shown. (a) shows points passing the \hs\ and \hb, whilst (b) shows points passing the \nt\ Higgs signal rate constraints}
    \label{fig:phenoRC2BR}
\end{figure}
Since we are interested in the product of the reduced coupling and $BR$, it is useful to plots their correlations.
The gluon-gluon higgs reduced coupling $g_{ggh}^2$ is shown in Fig.~\ref{fig:phenoRC2BR} plotted against $BR(h \to 2\aone$) for all the above assignments.
Overlaid are contours of constant $g_{ggh}^2 \times BR(h_i \to \aone\aone)$.
Two version of this plot have been made: one (Fig~\ref{fig:phenoRC2BRHSHB}) for points passing the \hs\ and \hb\ constraints, ignoring the \nt\ \chisq\ constraints; and Fig.~\ref{fig:phenoRC2BRNT} for points passing the \nt\ \chisq\ constraints ignoring the \hs\ and \hb\ constraints.
There are several important features to discern.
Generally, points where the heavier $h$ in the decay chain is the \hdisc-like object (blue and red) provide the largest $g_{ggh}^2 \times BR$ product, $\lesssim 0.2$ -- $0.5$, and therefore potentially the largest total $\sigma \times BR$.
These points have a very SM-like $ggh$ coupling as a result of meeting visible $ZZ/\diphoton/\bbbar$ signal rates, and are limited entirely by the experimental constraints on $BR(h \to \aone\aone)$.
Points where the heavier $h$ in the decay chain is \emph{not} the 125-like object have the opposite trend.
Given the lack of any other observed Higgs boson, these must have a small $ggh$ coupling, but are free to have sizeable $BR(h \to \aone\aone)$.
However their overall product is typically smaller, $\lesssim 0.05$.

A noticeable difference between the two plots is the allowed $BR(h \to \aone\aone)$, particularly in the $\hone = \hdisc$ scenario where \hs\ + \hb\ allows $BR \lesssim 0.5$, whilst \nt\ constraints this more severely to $BR \lesssim 0.2$.
Note that the aforementioned combined result from CMS and ATLAS falls halfway between these two values.
This is due to the differences between the programs: the experimental results they choose to use, and the manner in which they apply those results.
\begin{figure}[!htbp]
    \centering
        \includegraphics[width=0.5\textwidth]{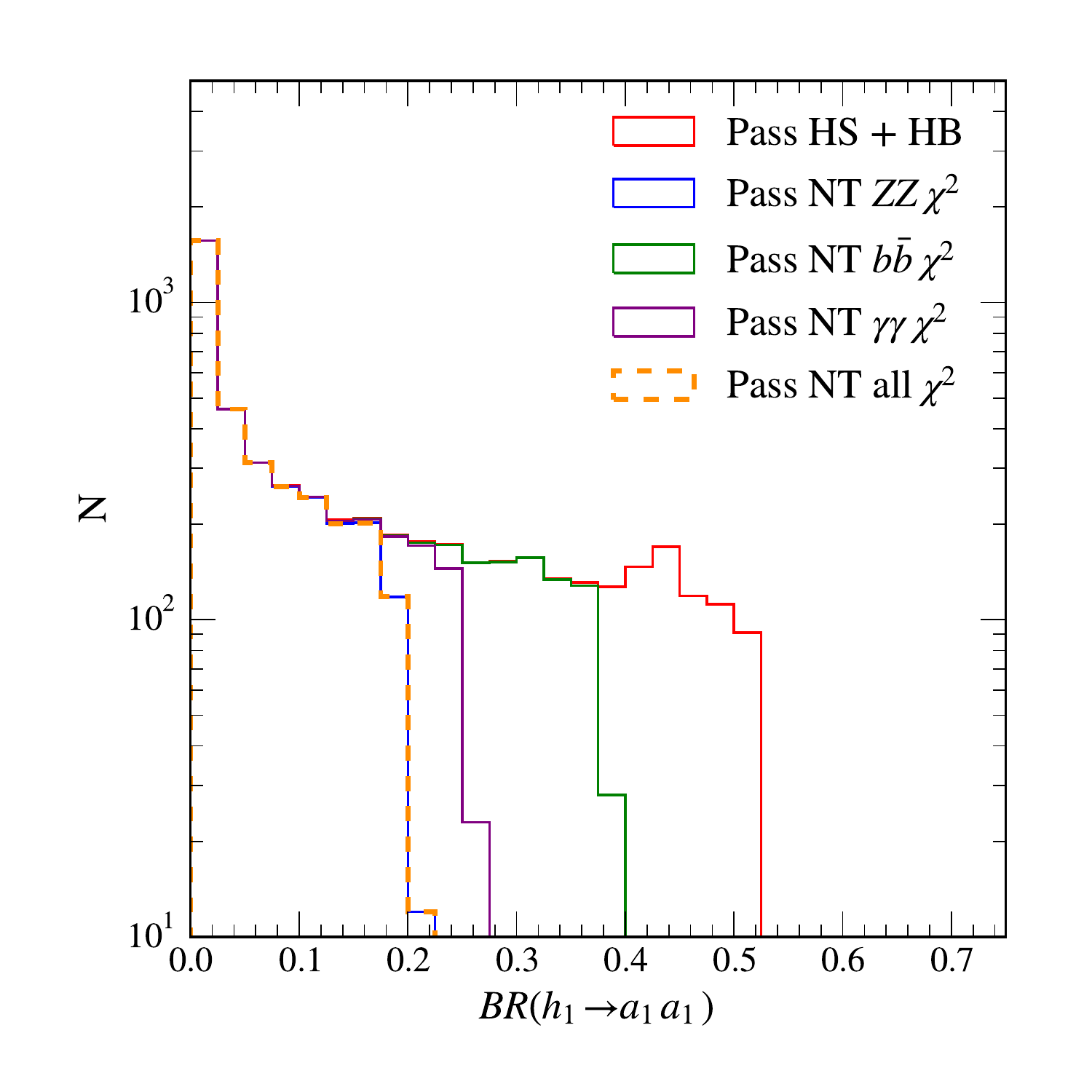}
    \caption{Distributions of $BR(\hone \to \aone\aone)$ for scenarios where $\hdisc = \hone$, comparing distributions passing individual channel signal rate \chisq\ constraints in \nt\ (blue, green, and purple), all three signal \chisq\ constraints in \nt\ (dashed orange), and points passing \hs\ and \hb\ (red). All points pass all other non-Higgs signal rate constraints such as flavour constraints (with relaxed constraints on \Delamu\ and \dmrelic\ as noted in the text), and have $\maone < 10.5\ \GeV$.}
    \label{fig:phenoBRH1NTHS}
\end{figure}
Fig.~\ref{fig:phenoBRH1NTHS} shows $BR(\hone \to \aone\aone)$, comparing distributions for models passing the \hs, for models passing each \nt\ \chisq\ constraint individually, and models passing all \nt\ \chisq\ constraints.
\nt\ performs a best-fit to each of the $ZZ/\diphoton/\bbbar$ final states as described in~\cite{Belanger:2013xza}, and compares the model compatibility by calculating \chisq\ for each final state.
Therefore if at least one of those fails, the point will be rejected.
We find that the $ZZ$ \chisq\ constraint places the strongest constraint on $BR(h \to \aone\aone)$.
However, \nt\ does not use information from other channels, such as \tautau.
\hs\ in contrast uses information from a much larger set of analyses (85 in version 1.4.0), and performs a global \chisq\ fit.
Therefore, one can have a large rate in a certain channel if it is compensated by a low rate in another channel.

\begin{figure}[!htbp]
    \centering
    \includegraphics[width=.7\textwidth]{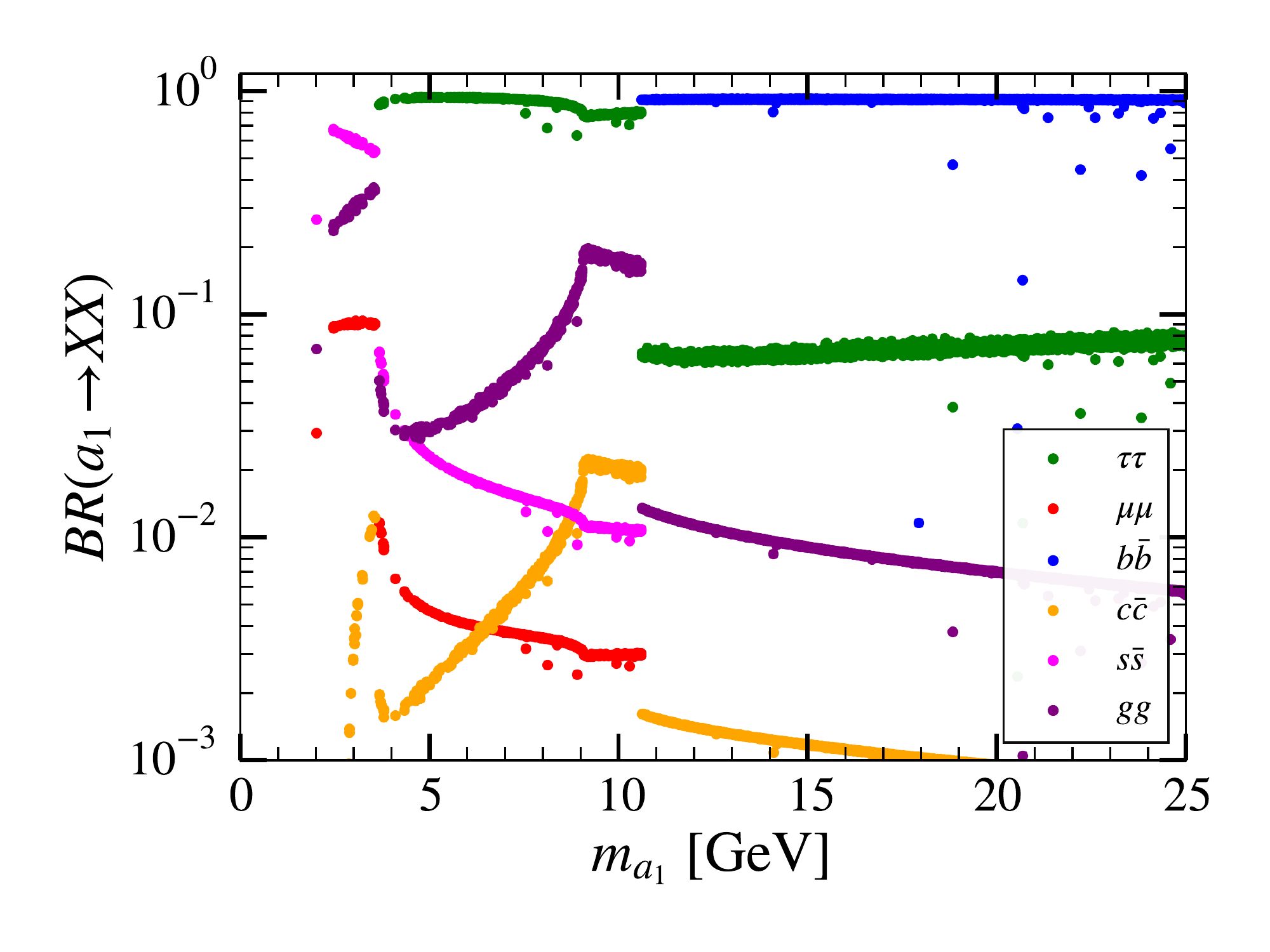}
    \caption{$BR(\aone \to 2X)$ as a function of \maone. All points here pass all the described constraints.}
    \label{fig:phenoBRa}
\end{figure}

The last piece of the eq. \ref{eq:split}, $BR(\aone \to 2X)$, is shown in Fig.~\ref{fig:phenoBRa}.
For each final state, at a given \aone, there is little variance over the range of input parameters.
This is a consequence of all decays depending on the same Yukawa couplings; the total width of the \aone\ can vary depending on its mixing and the value of \tanbeta, but in the branching ratios all this is factored out and we are left with functions of a few parameters in the Higgs sector that are already fixed by phenomenology.
The few points that deviate from the lines in Fig.~\ref{fig:phenoBRa} can be understood from the occasional presence of other channels, e.g.\ $\aone\to \gamma\gamma$ that is sometimes enhanced by large chargino loops.

Since the branching ratios are dependent on the Yukawa couplings, we see that it is the heaviest decay products that dominate and this manifests in the boundaries at $(\sim 3.5, 10.5\ \GeV)$ where heavier final states ($\tautau, \bbbar$) become kinematically viable, above which they become the favoured decay channels.
Note that for \tautau\ this threshold happens at $2m_\tau$ as expected, while for \bbbar\ the threshold is set to twice the B meson mass, which is somewhat larger than twice the $b$-quark mass (in principle there could be decays including mesons with $b$ quarks also just below this limit, but the calculation of such channels is very challenging and not included in \nt).

One striking feature of Fig.~\ref{fig:phenoBRa} is the behavior of $BR(\aone\to gg)$, which in the mass window 3.5~--~10.5~GeV is dominated by the $b$-quark loop.
The contribution from this loop increases rapidly until \maone\ reaches $\sim 9$~GeV at which point the quarks in the loop become real, after which it slowly decreases (due to increasing virtuality of the quarks).
This threshold does not coincide with the onset of the \bbbar\ channel since the loop behaviour is governed by the $b$-quark pole mass, and not the B meson mass that governs the threshold.
This behavior is replicated in $BR(\aone\to \ccbar)$ due to this channel being dominated by $\aone\to gg^*\to g\ccbar$, where $g^*$ is a virtual gluon.
The kink in the $BR(\aone\to gg)$ line at 9~GeV is also mirrored in the other branching ratios since a decreasing width to gluons will result in an increasing branching ratio for all other final states.

We further know that the width of a channel typically increases quickly with the mass of the mother particle just above its kinematic threshold, then increases slower when the phase space factors become less dominant.
This explains why, for example, $BR(\aone\to \tautau)$ increases in the region from 2\mtau\ to around 6~GeV, and in turn explains the decrease in $BR(\aone\to \mumu)$ and $BR(\aone\to \ssbar)$ in the same region.

Below 2\mtau, \ssbar\ is the dominant decay channel due to its relatively large mass, as well as colour factors that favour quarks over \mumu.
$BR(\aone\to \ssbar)$ decreases due to the increasing gluon final state, while $BR(\aone\to \mumu)$ stays constant as the tendency to decrease due to increasing $BR(\aone\to gg)$ is compensated by the fast increase in width due to being close to threshold.
There are also QCD effects giving quark channels a flatter curve close to threshold as compared to leptons; this is why \ssbar\ decreases while \mumu\ remains constant.
This is also why $BR(\aone \to \tautau)$ is increasing slightly above 10~GeV;  $BR(\aone \to \bbbar)$ increases somewhat slower than $BR(\aone \to \tautau)$ despite being closer to threshold.

\begin{figure}[!htbp]
    \centering
    \begin{subfigure}[t]{0.8\textwidth}
        \includegraphics[width=\textwidth]{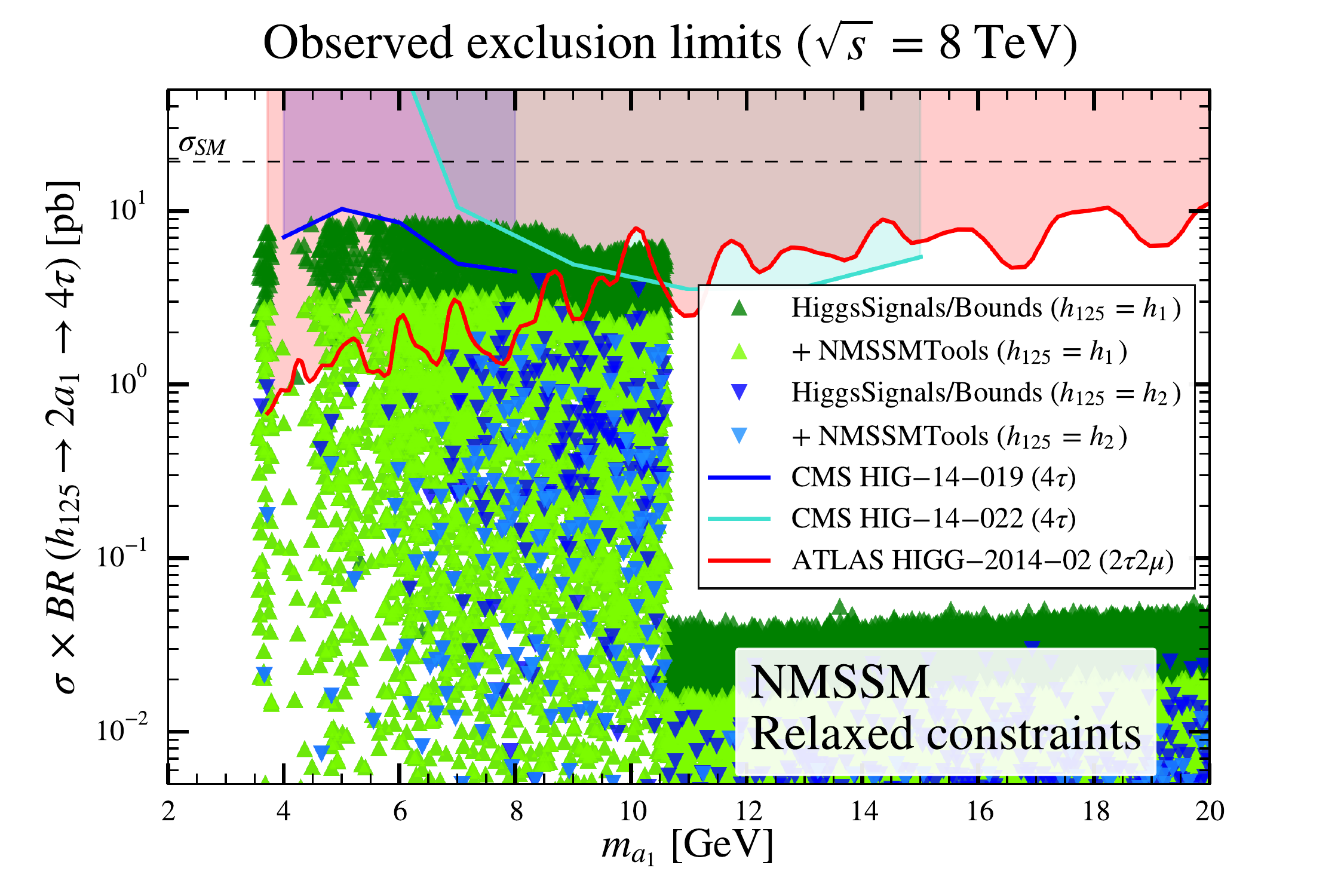}
        \caption{}
        \label{fig:phenoTotalXsec4tauLowerMass}
    \end{subfigure}
    ~
    \begin{subfigure}[t]{0.8\textwidth}
        \includegraphics[width=\textwidth]{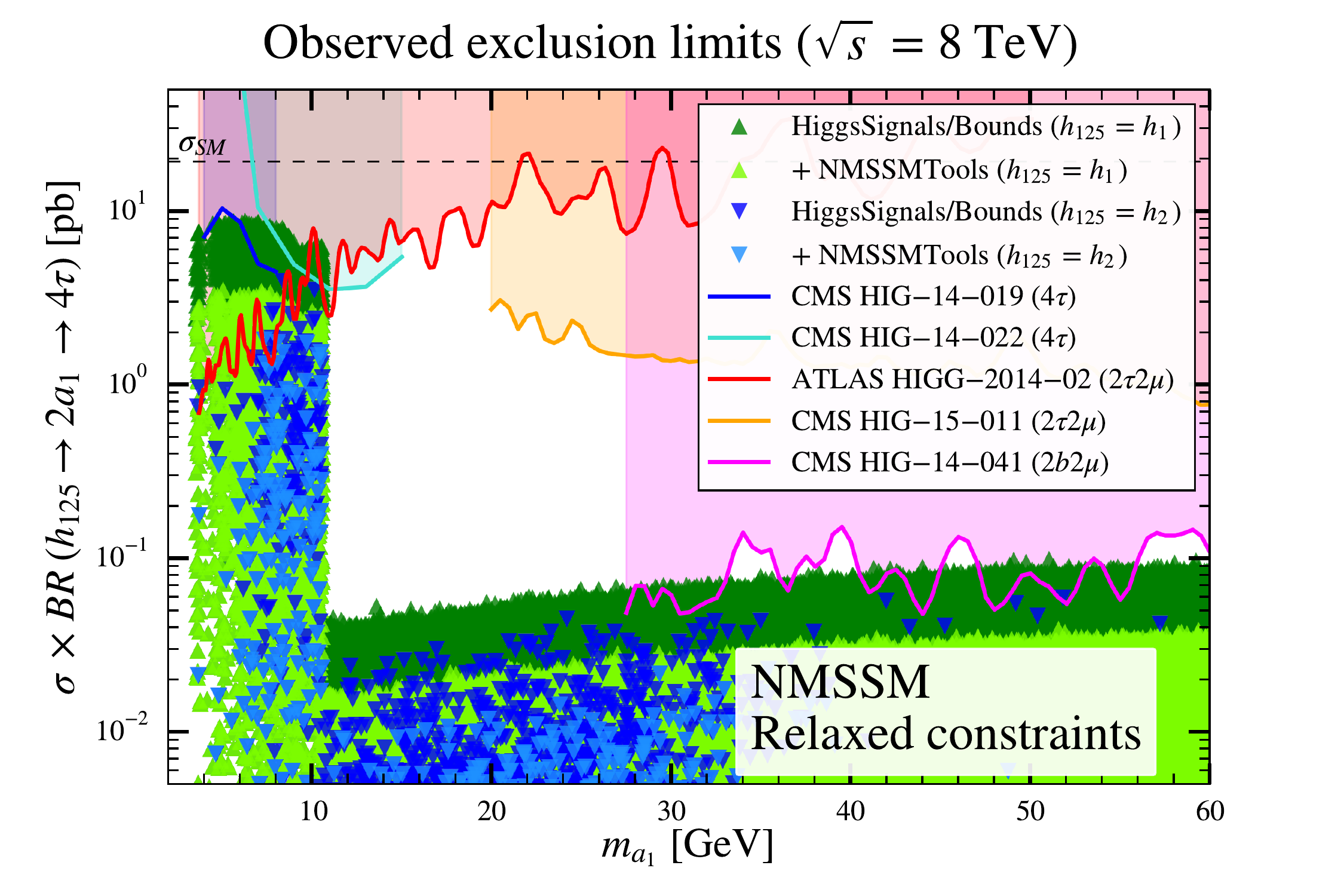}
        \caption{}
        \label{fig:phenoTotalXsec4tauHigherMass}
    \end{subfigure}
    \caption{Plots of $\sigma \times BR(gg \to h_i \to 2\aone \to 4\tau)$ versus \maone\ for various Higgs assignments in the NMSSM. Green upwards triangles are those where the heavier Higgs in the decay chain $h_i = \hone$, and blue downwards triangles are those where the heavier Higgs $h_i = \htwo$. Dark green/blue points are only required to satisfy Higgs rate constraints from \hs, whilst lighter green/blue points must also pass \nt\ Higgs rate constraints. All points pass a ``relaxed'' set of constraints, where we also require all other \nt\ constraints, but allow any $\Delamu > 0$, $\dmrelic < 0.131$, and ignore limits on \rd\ and \rds. Overlaid are observed exclusion regions from the relevant analyses. The SM cross-section at $\sqrt{s} = 8\ \TeV = 19.27\ \pb$ is also shown for reference. The top plot focuses on the low mass region. }
    \label{fig:phenoTotalXsec4tau}
\end{figure}
From the above studies, we expect total \fourtau\ cross-sections up to $\sim 19.3 \times 0.2 \times 0.9^2 \simeq 3\ \pb$ if one applies the \nt\ Higgs signal rate constraints, or even up to $\sim 8\ \pb$ if one uses the \hs\ constraints.
The experimental Higgs signal rate measurements are therefore the limiting factor in determining the total cross-section due to their impact on $BR(h \to 2\aone)$, and not any particular model feature.
We now combine all these pieces together, and plot the total cross-section as a function of $\maone$.
We start by considering the \fourtau\ final state in the NMSSM.
This is shown in Fig.~\ref{fig:phenoTotalXsec4tau}, where $\sigma \times BR(gg \to h_i \to 2\aone \to 4\tau)$ has been plotted against \maone, for masses greater than 2\mtau, with different assignments for \hdisc, and different Higgs signal rate requirements applied.
There are very few $\htwo \to 2\hone \to 4\tau$ points, and these have not been shown  due to significantly smaller cross-sections.
Points are required to pass the ``relaxed'' set of constraints, where we also require all other \nt\ constraints but allow any $\Delamu > 0$, $\dmrelic < 0.131$, and ignore limits on \rd\ and \rds.
Requiring lower bounds on \Delamu\ and \dmrelic\ does not change the overall result, and only reduces the overall number of points.
Overlaid are the observed exclusion limits from relevant searches.
One can see a wide variety of predicted cross-sections compatible with current experimental constraints, ranging from $< 1\ \fb$ up to $8\ \pb$.
As previously mentioned, models with $h_i = h_1$ (of which many have $\hone = \hdisc$) generally have a larger cross-section than those with $h_i = h_2$.
The large decrease in cross-section for masses $\maone > 2\mb \sim 10.5\ \GeV$ is due to the decrease in $BR(\aone\to\tau\tau)$ as the \bbbar\ final state becomes kinematically available.
The ATLAS \twotautwomu\ analysis is more powerful for masses 4 -- 10 GeV, especially at smaller masses, and is therefore complementary to the \fourtau\ analyses which lose sensitivity at smaller masses.
This analysis can exclude a significant number of points of $\hone \to 2\aone$, excluding cross-sections as small as 1~--~2~\pb, even taking into account the more restrictive Higgs signal rate constraints from \nt.
However, it is not yet sensitive enough to probe the alternate scenario where $\htwo \to 2\aone$.
The \fourtau\ analyses start to intrude on the model space, although only if one assumes the more relaxed rate constraint from \hs.
These excluded points are typically those where $h_i = \hdisc$, as such configurations often give a larger cross-section as shown in Fig.~\ref{fig:phenoRC2BR}.

A minor detail seen in Fig.~\ref{fig:phenoTotalXsec4tauLowerMass} is that the rates can go slightly higher when $\hone=\hdisc$.
This is a somewhat complicated effect from the structure of the parameter space; first, if \lam\ is large it is difficult to achieve acceptable SM signal rates for \hdisc\ if $\maone<\mh/2$, mostly because $BR(\hdisc\to\aone\aone)$ tends to increase with \lam, but also due to interplay with \lam\ affecting the mixing of the \hdisc.
Furthermore, if \lam\ is not too large we can only have $\htwo=\hdisc$ if \kap\ is also small (the singlet scalar mass goes as $\kap s$ and since $\lam s$ cannot be too small $\kap > \lam$ means the singlet scalar is heavy).
The coupling $\hdisc\aone\aone$ has a term proportional to $\lam\kap$ which can saturate $BR(\hdisc\to\aone\aone)$ with respect to the \hdisc\ signal rates if \kap\ is large. Hence the rate shown in Fig.~\ref{fig:phenoTotalXsec4tauLowerMass} can reach its maximum for $\hone=\hdisc$ but struggles to do so for $\htwo=\hdisc$.

Expanding our mass range up to 60 GeV means the limits from the CMS \twotautwomu\ and \twobtwomu\ analyses can also be included.
This is shown in Fig.~\ref{fig:phenoTotalXsec4tauHigherMass}, where one can see that the latter analysis is powerful enough to start to probe phase space (if one uses the more optimistic constraints from \hs).
The \twotautwomu\ analysis is not yet able to probe NMSSM phase space.
However it could offer some sensitivity if one were instead dealing with a model where $\aone \to \tau\tau$ was enhanced over $\aone \to \bbbar$, for example a Type III (IV) 2HDM with large (small) \tanbeta.
Crucially direct searches have similar, often better, sensitivity measuring $BR(\hdisc \to \mathrm{BSM})$ than limits from indirect searches, assuming BSM is solely \aone\aone.

One additional point to note in this Figure is the lack of points with $\maone \sim 4$ -- $4.5\ \GeV$ and $\maone \sim 5$ -- $5.5\ \GeV$.
These masses are heavily suppressed due to flavour constraints: the former mass range is excluded by $BR(B \to X_s \mu\mu)$, whilst the latter range is excluded by $B_{s,d} \to \mu\mu$.

\begin{figure}[!htbp]
    \centering
    \includegraphics[width=0.8\textwidth]{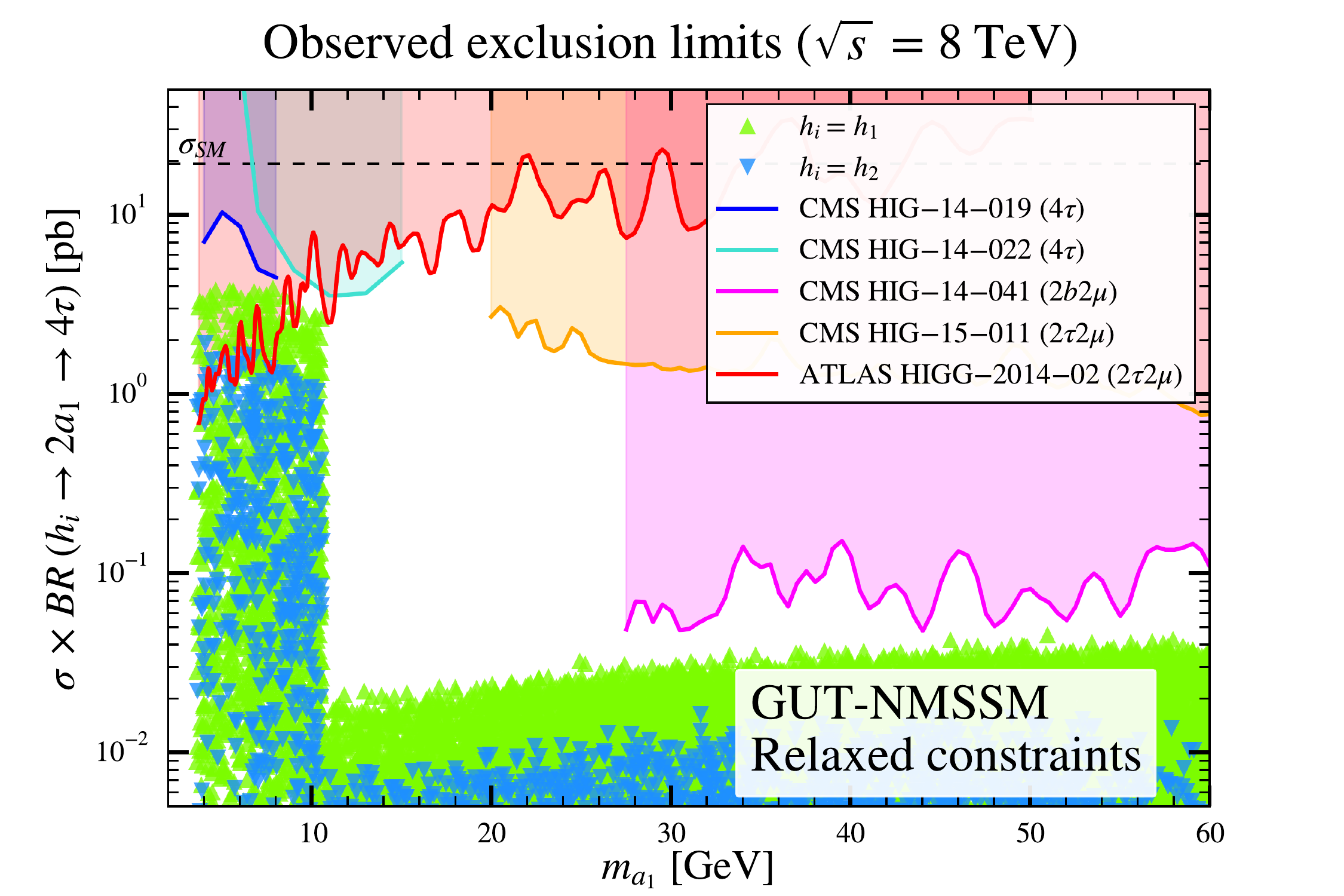}
    \caption{Plots of $\sigma \times BR(gg \to h_i \to 2\aone \to 4\tau)$ versus \maone\ for various Higgs assignments in the GUT-constrained NMSSM. Green upwards triangles are those where the heavier Higgs in the decay chain $h_i = \hone$, and blue downwards triangles are those where the heavier Higgs $h_i = \htwo$.}
    \label{fig:pheonTtoalXsec4tauGUT}
\end{figure}
One might also consider the cross-section as predicted in the GUT-constrained NMSSM, shown in Fig.~\ref{fig:pheonTtoalXsec4tauGUT} for the \fourtau\ final state.
This shows a very similar result to that in Fig.~\ref{fig:phenoTotalXsec4tauHigherMass}, whilst the limiting factor remains that on Higgs-to-Higgs decays.
This bound is in general easy to satisfy, and hence the upper limit on possible rates in the channels we have studied is essentially independent of model details such as GUT scale unification.

\begin{figure}[!htbp]
    \centering
    \includegraphics[width=0.8\textwidth]{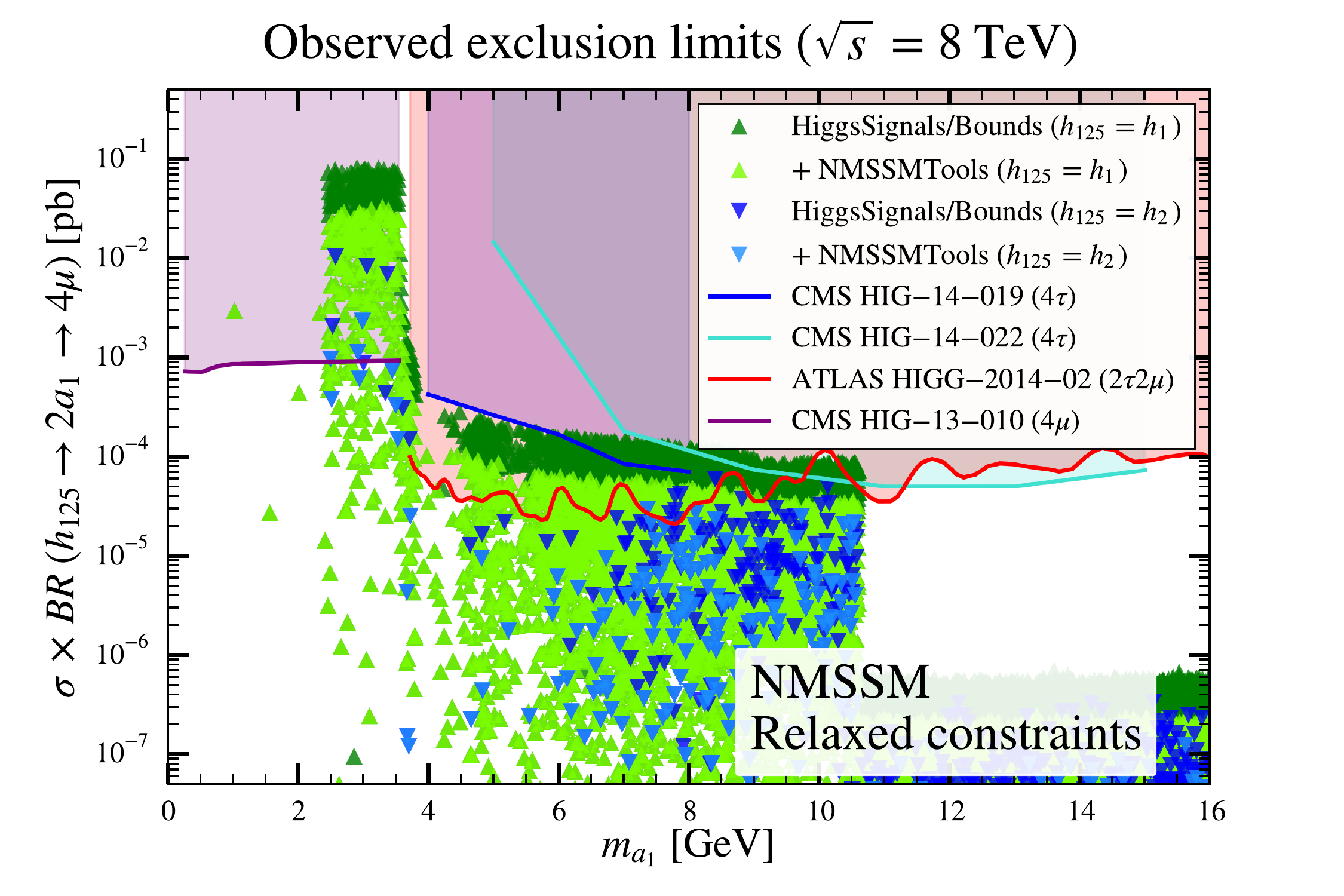}
    \caption{Plot of possible $\sigma \times BR(gg \to h_i \to 2\aone \to 4\mu)$ against \maone for various Higgs assignments and constraints in the NMSSM. For details see the caption of Fig.~\ref{fig:phenoTotalXsec4tau}. Overlaid are observed exclusion regions from \fourtau, \twotautwomu, and \fourmu\ analyses.}
    \label{fig:phenoTotalXsec4mu}
\end{figure}

Let us also consider the \fourmu\ final state.
$\sigma \times BR(gg \to h_i \to 2\aone \to 4\mu)$ in shown in Fig.~\ref{fig:phenoTotalXsec4mu} as a function of \maone.
Colours and shape assignments are the same as for the \fourtau\ figure.
The relevant experimental limits now include the CMS \fourmu\ search.
This probes cross-sections down to 1~\fb, and therefore excludes many model points.
There are almost no points below $\maone < 2.5\ \GeV$.
Points with $\maone < 1\ \GeV$ are rejected on grounds that their decay widths are difficult to calculate accurately due to hadronisation effects and QCD effects, while points with $1 < \maone < 2.5\ \GeV$ are rejected by constraints on $B \to X_s \mu\mu$.

\begin{figure}[!htbp]
    \centering
    \begin{subfigure}[t]{0.44\textwidth}
        \includegraphics[width=\textwidth]{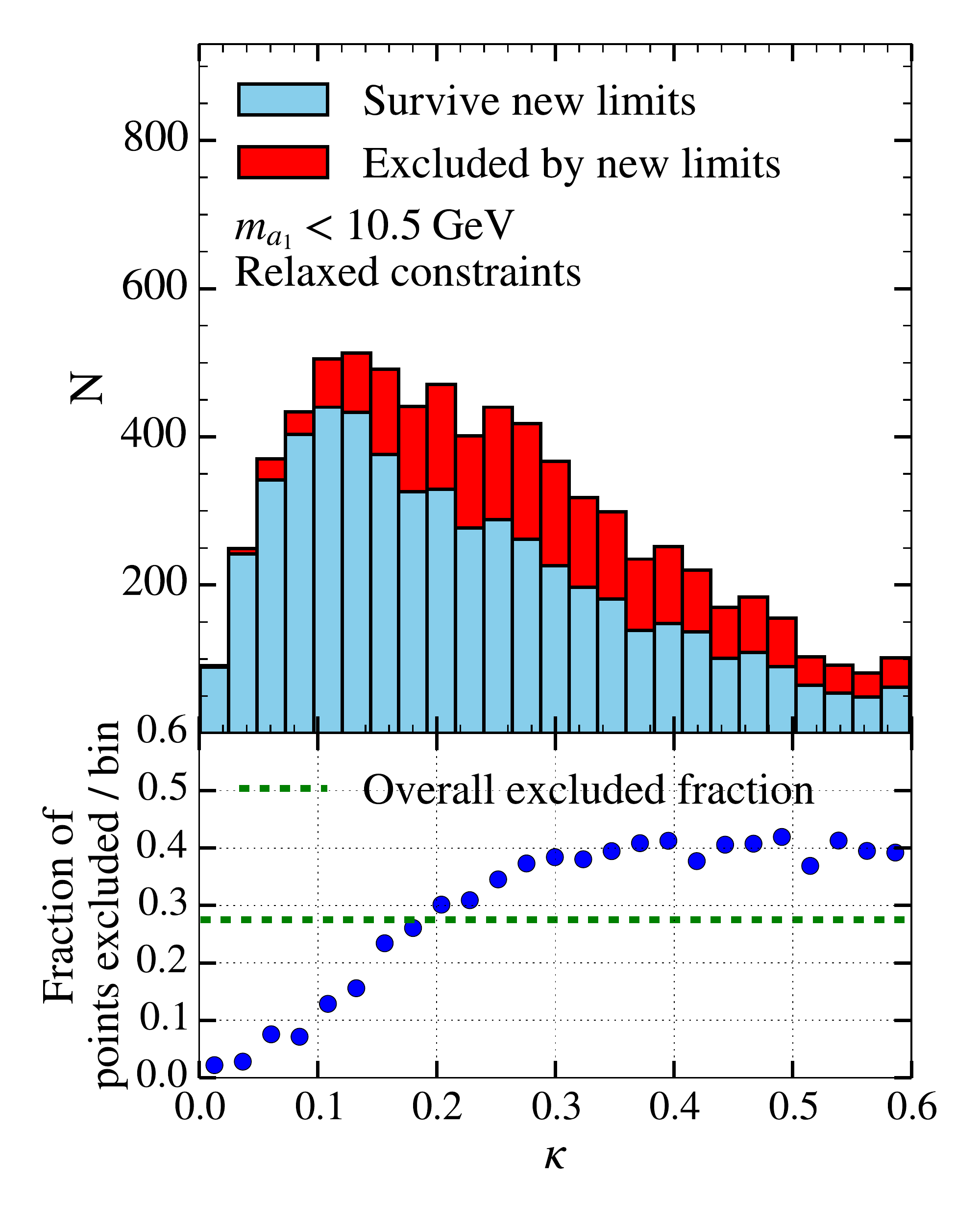}
        \caption{\kap}
        \label{fig:phenoConstraintHistsKappa}
    \end{subfigure}
    \begin{subfigure}[t]{0.44\textwidth}
        \includegraphics[width=\textwidth]{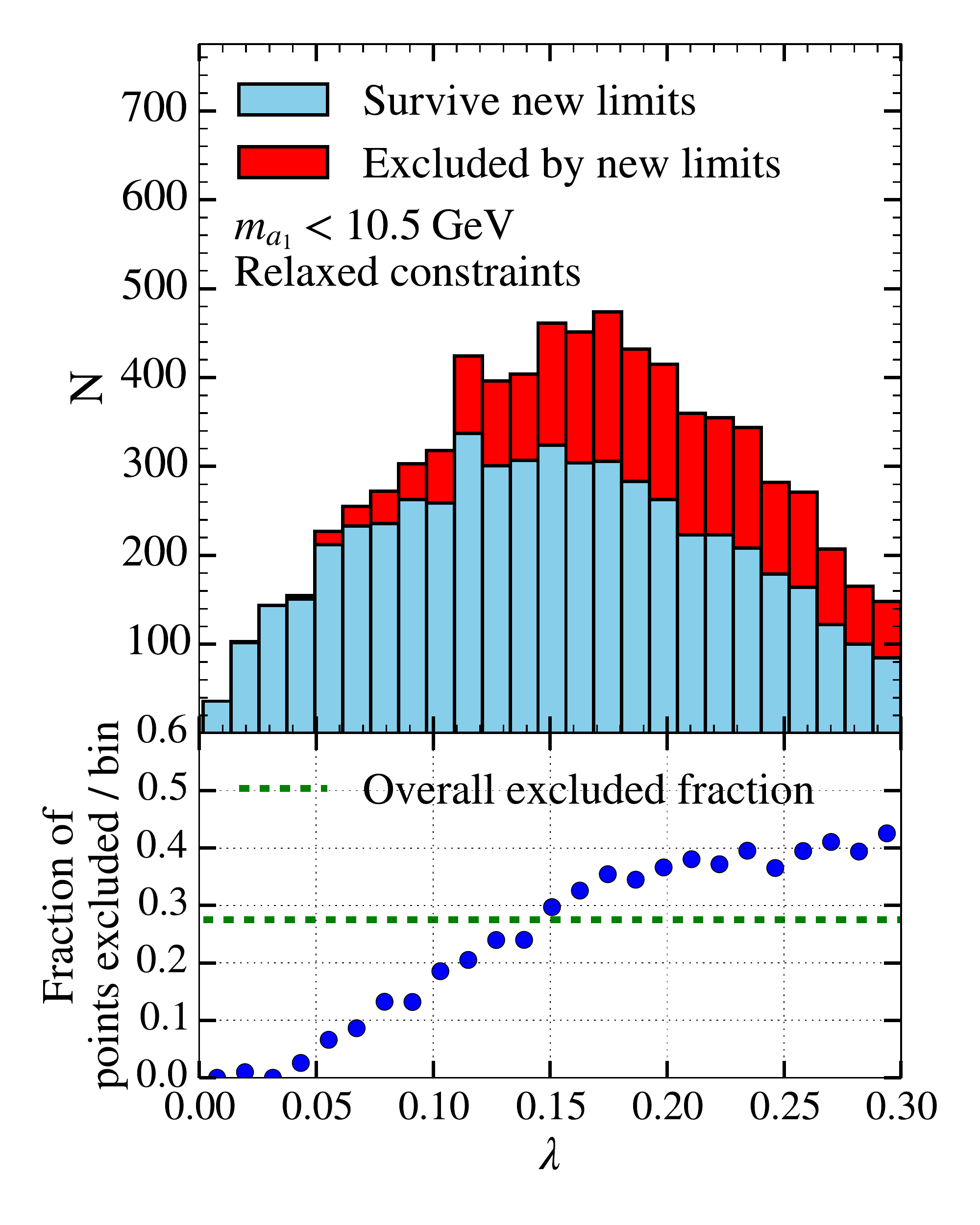}
        \caption{\lam}
        \label{fig:phenoConstraintHistsLambda}
    \end{subfigure}
    \begin{subfigure}[t]{0.44\textwidth}
        \includegraphics[width=\textwidth]{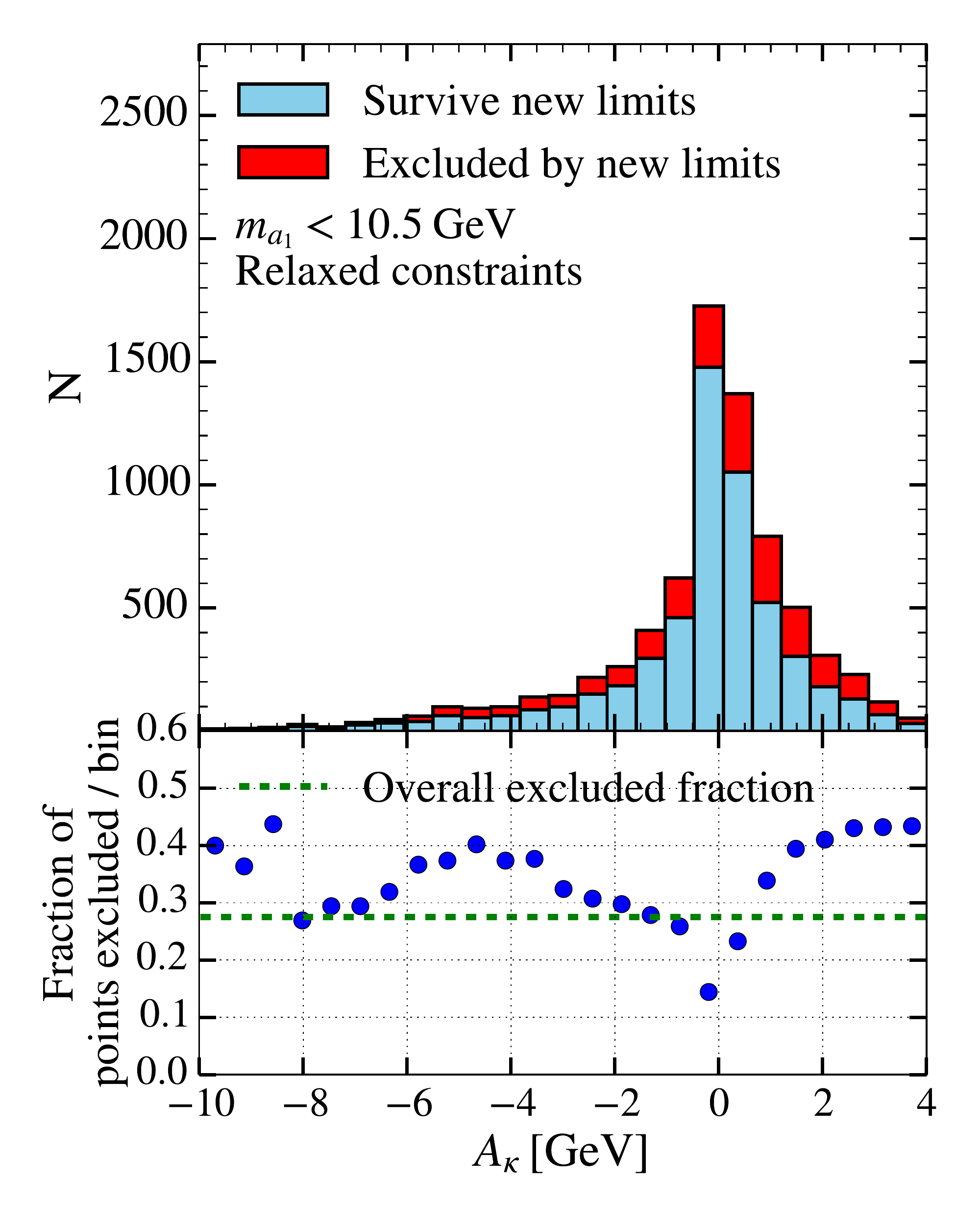}
        \caption{\akap}
        \label{fig:phenoConstraintHistsAKappa}
    \end{subfigure}
    \caption{Distributions of input parameters in the NMSSM, divided into points surviving the new limits (blue) and failing the new limits (red). Also shown is the fraction of points failing per bin, and the overall average fraction of points failing.}
    \label{fig:phenoConstraintHists}
\end{figure}
\begin{figure}[!htbp]
    \centering
    \includegraphics[width=\textwidth]{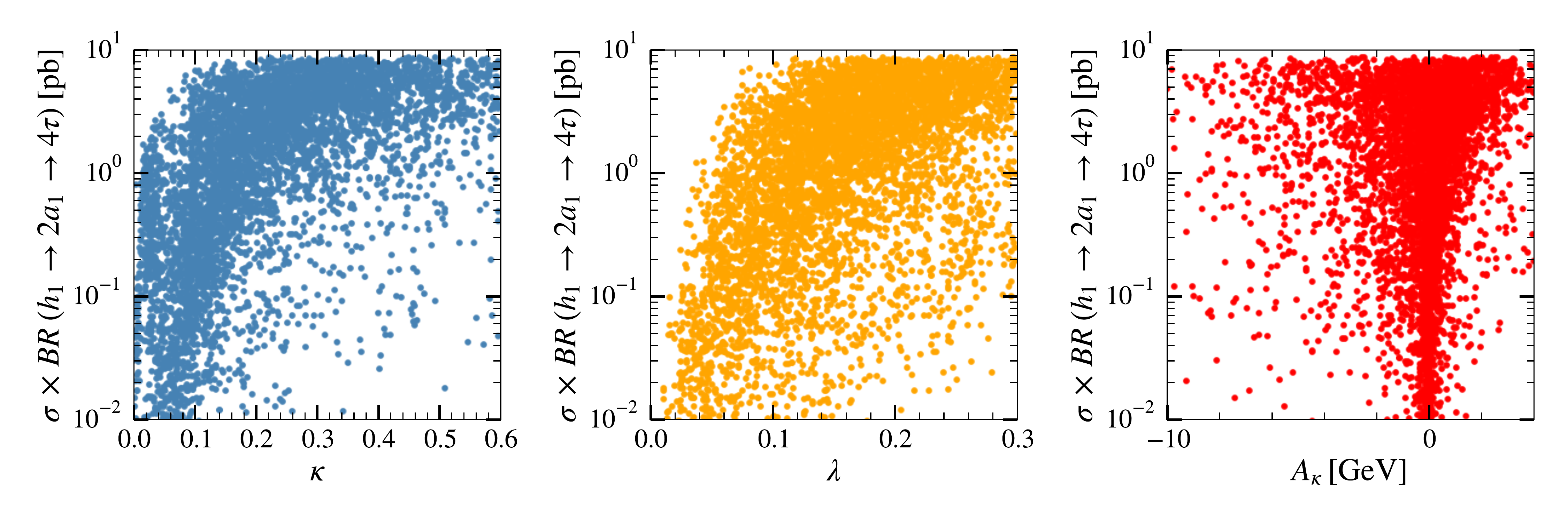}
    \caption{Plots of $\sigma \times BR(gg \to \hone \to 2\aone \to 4\tau)$ against (a) \kap, (b) \lam, and (c) \akap\ in the NMSSM. Points have $\maone < 10.5\ \GeV$, and pass \hs\ constraints as well as the other non-Higgs constraints.}
    \label{fig:phenoXsecInputParams}
\end{figure}

Since the total cross-section is driven by the limit on $BR(h_1 \to 2\aone)$, which in turn has a strong dependence on several input parameters (Fig.~\ref{fig:phenoBrh1a1a1Hist2D}), one can look at the impact of these new limits on possible model parameter values.
\kap, \lam, and \akap\ are of particular interest.
Histograms of the distributions are shown in Fig.~\ref{fig:phenoConstraintHists}, where they have been divided into points surviving all new constraints (blue) and failing any of the new constraints (red) for models with $\maone < 10.5\ \GeV$.
Also shown is the ratio of failing to surviving points for each bin, and the global fraction of points failing.
\kap\ and \lam\ show a clear trend that higher values are more likely to be excluded, which is expected as the $h_i\aone\aone$ coupling depends on $\lam^2$ and \lam\kap.
Additionally, small positive values of \akap\ also show a similar trend.
Although these new constraints do not place a hard limit on values of these parameters, as such limits improve over time they will point towards models with smaller values.
If experimental limits can exclude cross-sections down to 100s of \fb\ then Fig.~\ref{fig:phenoXsecInputParams} shows that these parameters may be far more constrained, particularly \kap\ due to the ``knee'' shape of its distribution, and \akap\ due its ``wedge'' shaped distribution.
Constraining \lam\ to smaller values is of particular interest since the tree-level Higgs mass has an additional contribution $\propto \lam$ compared to the MSSM, one of the strengths of the NMSSM with respect to the MSSM.
If this extra contribution is small, then a larger (and potentially more uncomfortable) degree of fine-tuning is required to achieve a mass of 125 \GeV.

We now consider the results of scans for the other models.
\begin{figure}[!htbp]
    \centering
    \includegraphics[width=0.8\textwidth]{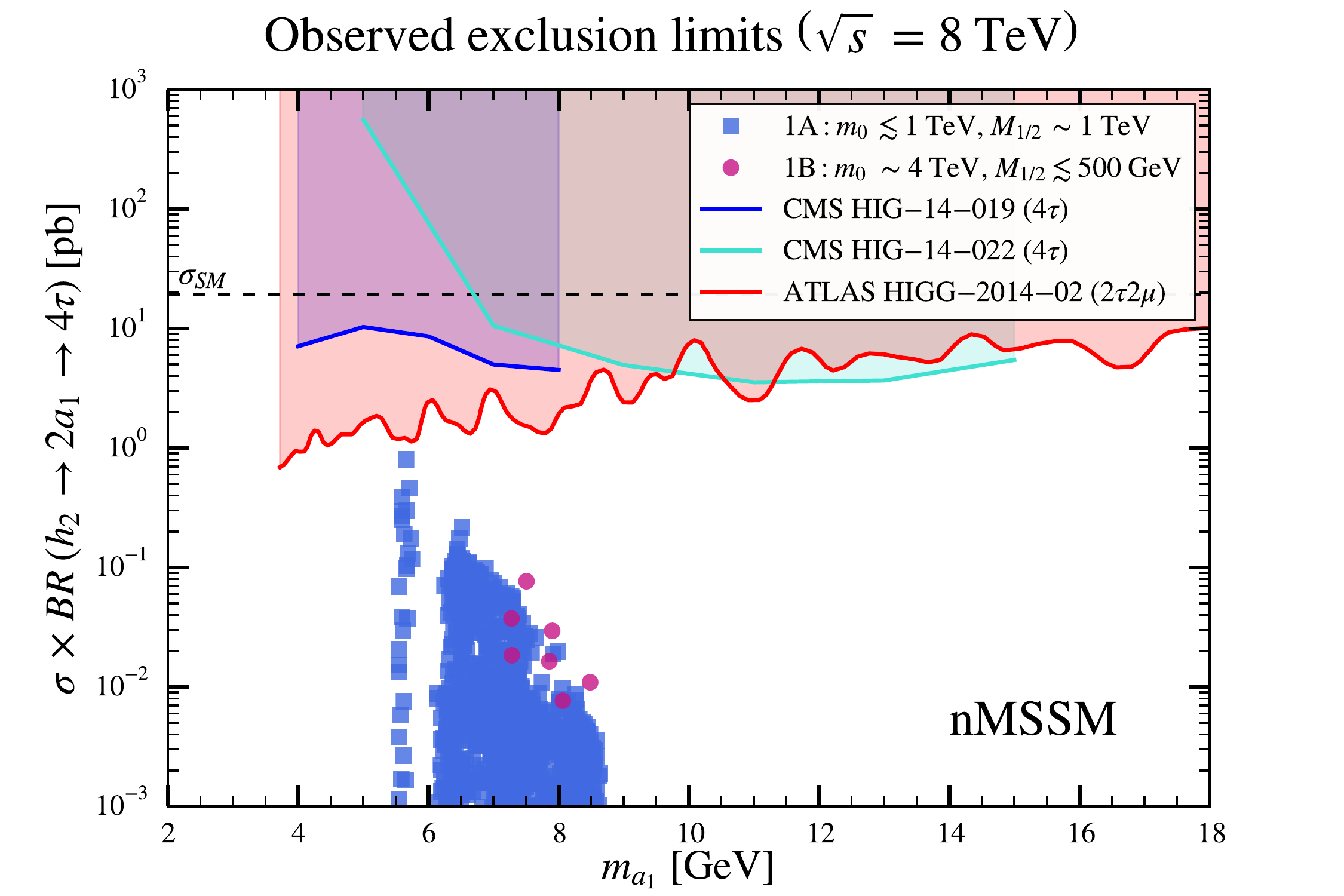}
    \caption{Plot of $\sigma \times BR(gg \to h_2 \to 2\aone \to 4\tau)$ versus \maone\ for various Higgs assignments in the nMSSM. Blue squares are in region 1A (small $m_0$ and $M_{1/2}$), whilst violet-red circles are in region 1B (larger $m_0$ and smaller $M_{1/2}$). Overlaid are observed exclusion regions from various analyses. The SM cross-section at $\sqrt{s} = 8\ \TeV = 19.27\ \pb$ is also shown for reference.}
    \label{fig:phenoLittlenMSSMscan}
\end{figure}
Although one might assume the nMSSM would give similar results to the NMSSM, the spectra of cross-sections and masses as shown in Fig.~\ref{fig:phenoLittlenMSSMscan} is very different.
From the Figure we observe that the $a_1$ mass (which we recall has a large singlino component) is constrained to be in a small mass window, between $\sim$5 and $\sim$11 GeV.
As mentioned, this is due to the fact that the DM candidate (the lightest neutralino $\tilde \chi^0_1$, which is almost a pure singlino) has a mass around 5 GeV and its relic abundance is fixed via annihilation through the lightest pseudoscalar $a_1$.
This constrains \maone\ to be near the resonant peak, with $0<m_{a_1}-2m_{\tilde \chi^0_1}<$ 1 GeV~\cite{Barducci:2015zna}.
The singlino nature of the lightest pseudoscalar and the small mass of $\tilde \chi^0_1$ also makes $BR(a_1\to\tilde \chi^0_1 \tilde \chi^0_1)$ to be the dominant decay channel for the lightest pseudoscalar, therefore causing a reduction of the $a_1\to\tau\tau$ rates and hence of the $4\tau$ cross sections.
Finally, we mention that in Fig.~\ref{fig:phenoLittlenMSSMscan} we have included all points surviving the scan of~\cite{Barducci:2015zna}. However the values of $m_0$ and $M_{1/2}$ have a strong impact on the particle spectrum of the model.
In particular, the region with small $M_{1/2}$ features a light gluino which is on the edge of the exclusion from 8 TeV searches that will soon be tested by the current run of the LHC.
A similar consideration can be made for the region with small $m_0$, that features light scalar superpartners (especially stops and sleptons).
In this respect, the results of Fig.~\ref{fig:phenoLittlenMSSMscan} has to be intended as to show only the current reach of light scalar searches in a different supersymmetric scenario, thus neglecting information arising from other LHC searches.

\begin{figure}[!htbp]
    \centering
    \begin{subfigure}[t]{0.8\textwidth}
        \includegraphics[width=\textwidth]{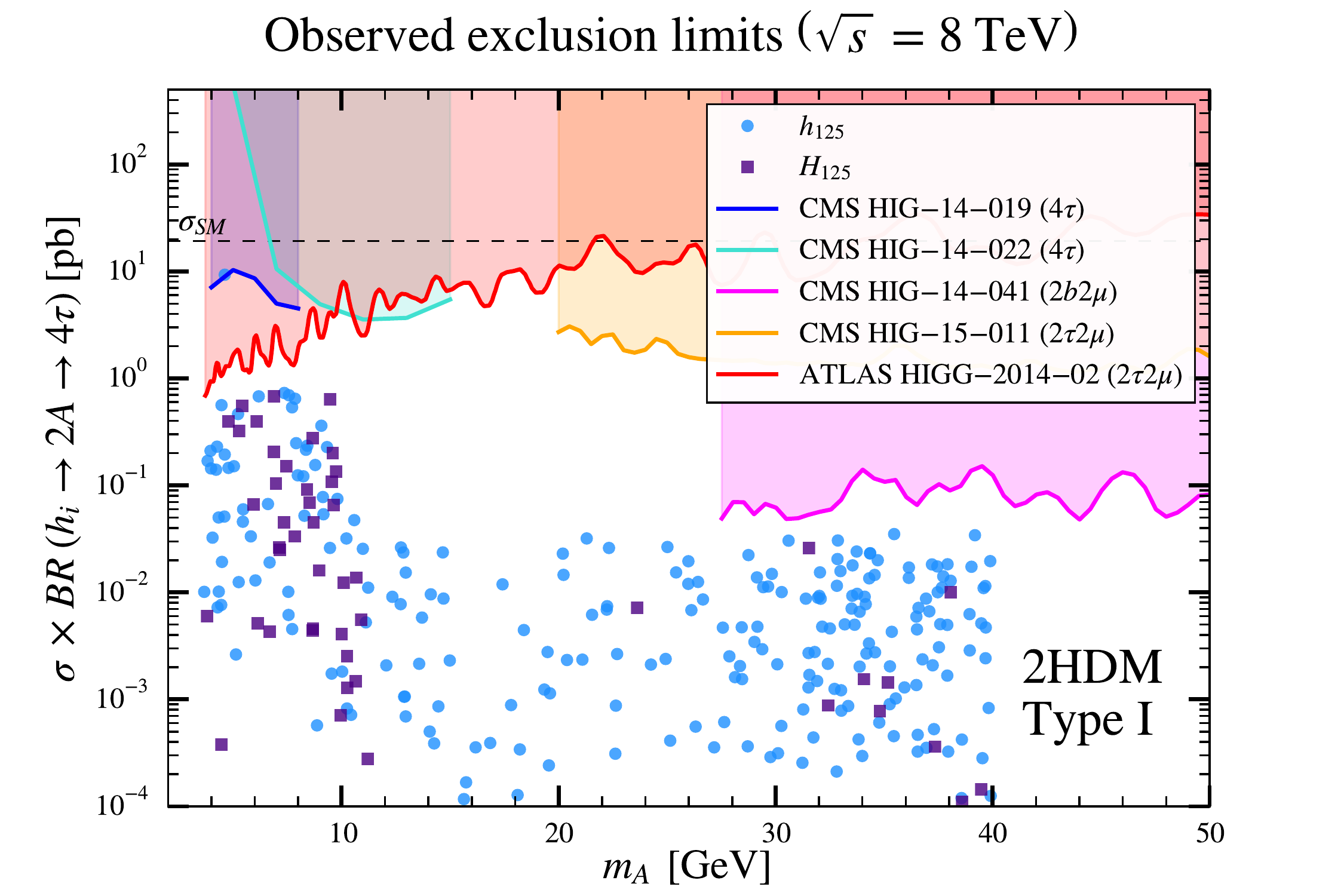}
        \label{fig:pheno2HDMscanTypeI}
    \end{subfigure}
    ~
    \begin{subfigure}[t]{0.8\textwidth}
        \includegraphics[width=\textwidth]{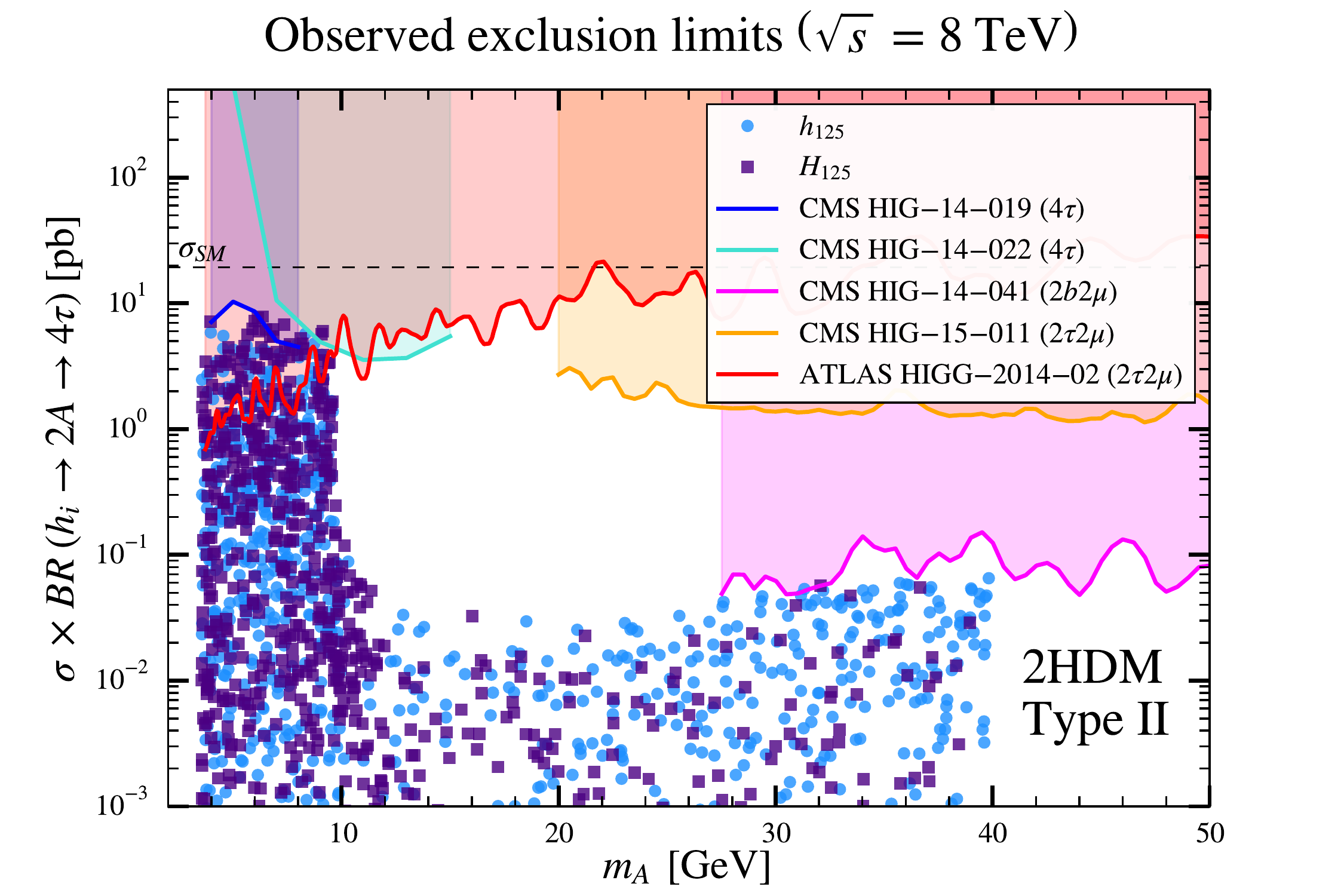}
        \label{fig:pheno2HDMscanTypeII}
    \end{subfigure}
    \caption{Plot of $\sigma \times BR(gg \to h/H \to 2A \to 4\tau)$ versus $m_A$ for different Higgs assignments in the Type I and Type II 2HDM. Blue circles are those where the lighter scalar is \hdisc, whilst purple squares are those where the heavier scalar is \hdisc. Overlaid are observed exclusion regions from various \fourtau\ and \twotautwomu\ analyses. The SM cross-section at $\sqrt{s} = 8\ \TeV = 19.27\ \pb$ is also shown for reference.}
    \label{fig:pheno2HDMscan}
\end{figure}

Lastly, we return to the more general Type I and II 2HDMs.
Shown in Fig.~\ref{fig:pheno2HDMscan} is the result of those scans for the \fourtau\ final state.
Both possible assignments for \hdisc\ are shown.
The Type II models predict significantly larger cross-sections ($\sim$ 7 -- 8 pb) than in the Type I ($\sim$ 1 pb), due to a different $\tan\beta$ dependence of the light pseudoscalar couplings, which favours higher $BR(a_1\to \tau\tau)$ in type II with respect to type I.
In the Type I model, there are also fewer points with $H = \hdisc$, whereas in the Type II model there is no such favouritism.
In the Type II model, the cross-section range is similar to that in the NMSSM, since the limiting factor is the experimental constraint on $BR(h/H \to AA)$. Overall, we find that in both configurations of 2HDM Yukawas, current searches targeting light (pseudo)scalars are starting to scratch the edge of the predicted models cross sections thus making the LHC Run 2 a crucial probe also for these scenarios.

\clearpage
\newpage
\section{Conclusion}
\label{sec:conclusion}

In summary, following the end of Run 1 at the LHC, we have assessed the status of direct searches for a light neutral Higgs boson in popular BSM scenarios with two Higgs doublets in both non-SUSY (2HDMs Type-I and II) and SUSY (NMSSM and nMSSM) frameworks, the latter also including
an additional Higgs singlet field.
The ability to extract signals of such a particle state would not only be a proof of a non-SM Higgs  sector but also a circumstantial evidence of either a non-minimal SUSY (as such a signal is not available in the MSSM) or a non-SUSY scenario. The mass region concerned is
up to 50 GeV or so. In such a range, the accessible decays, depending on the actual value of the light Higgs boson mass, are $\mu^+\mu^-$,
$\tau^+\tau^-$ and $b\bar b$. The topologies searched for exploit a cascade chain wherein such a light Higgs state is produced in pairs from the decay of another Higgs state, where the latter could be the SM-like Higgs boson discovered in 2012 at the LHC or not. Hence, final state topologies are a combination of two amongst the aforementioned two-particle decays. Those pursued experimentally during Run 1, covering the
discussed mass interval, were $4\tau$,  $2\tau 2\mu$, $4\mu$ and $ 2b2\mu$. We exploited public results produced by ATLAS and CMS for these final states in order to set limits on the parameter space of all four scenarios considered, 2HDMs Type-I and II plus NMSSM and nMSSM.
In doing so we have employed different numerical tools implementing these theoretical scenarios and/or corresponding
experimental constraints, so as to enable us to distinguish  genuine physics differences in the scope afforded by the various channels
from artifacts due to the different degrees of accuracy in the  model implementation.

Needless to say, the yield of these channels is not currently available in public tools, nor is the dedicated recasting procedure from one signature to another and onto a particular theoretical model that we have pursued here, so that our study represents an advancement in relation to current phenomenological knowledge, as the latter primarily rely on the study of SM-like signatures of additional Higgs states. Specifically, we have established that combinations of such signatures exclude substantial regions of the 2HDM Type-II (typically for masses below 10 GeV) but not in Type-I, which remains essentially untouched. As for the NMSSM and nMSSM, again, only one of these two scenarios is currently probed over significant portions of its parameter space (NMSSM), over the same mass range, while the other (nMSSM) is largely unaffected. Furthermore, the experimental searches considered do not make any assumption on the nature (whether scalar or pseudoscalar) of the light Higgs states, hence our results are applicable to whichever Higgs-to-two-Higgs decay pattern. We finally remark that all available experimental constraints were implemented, stemming from collider searches, both past (from LEP/SLC and Tevatron) and current (from LHC Run 1 and 2) ones, as well as from flavour and DM probes.

An obvious outlook of our work is to extend our  analysis to forthcoming LHC Run 2 results for these and similar topologies, wherein we expect a substantially increased experimental sensitivity to the theoretical models considered.

\section{Acknowledgements}
\label{sec:acknowledgements}
RCA and SM are supported in part through the NExT Institute.
The use of the CIS computer cluster at NCBJ is gratefully acknowledged.
SM acknowledges partial support from the STFC Consolidated Grant ST/J000396/1.
The use of the DICE computer cluster at UoB is also gratefully acknowledged.
Thanks to Tim Stefaniak for instructions on how to modify \hs.
Thanks to Ulrich Ellwanger for discussions on Fig.~\ref{fig:phenoBRa}.

\clearpage
\newpage
\bibliography{refs.bib}
\bibliographystyle{JHEP}

\end{document}